\documentclass[
    ,final            % use final for the camera ready runs
  ,numberedheadings % uncomment this option for numbered sections
  ]
 {aipproc}

\layoutstyle{6x9}

\newcommand{\lsim}{\raisebox{-4pt}{$\,\stackrel{\textstyle <}{\sim}\,$}}
\newcommand{\gsim}{\raisebox{-4pt}{$\,\stackrel{\textstyle >}{\sim}\,$}}

\begin{document}

\title{Dilepton Spectroscopy of QCD Matter at Collider Energies}

\classification{}
\keywords{Medium Modifications of Hadrons, Chiral Symmetry Restoration, Thermal 
Dilepton Emission, Ultrarelativistic Heavy-Ion Collisions}

\author{Ralf Rapp}{
  address={Cyclotron Institute and Department of Physics \& Astronomy, 
Texas A\&M University, College Station, TX 77843-3366, USA}
}

\begin{abstract}
Low-mass dilepton spectra as measured in high-energy heavy-ion collisions are a 
unique tool to obtain spectroscopic information about the strongly interacting 
medium produced in these reactions. Specifically, in-medium modifications of the 
vector spectral function, which is well known in the vacuum, can be deduced from 
the thermal radiation off the expanding QCD fireball. This, in particular, allows 
to investigate the fate of the $\rho$ resonance in the dense medium, and possibly 
infer from it signatures of the (partial) restoration of chiral symmetry, which 
is spontaneously broken in the QCD vacuum. After briefly reviewing calculations of 
thermal dilepton emission rates from hot QCD matter, utilizing effective hadronic 
theory, lattice QCD or resummed perturbative QCD, we focus on applications to 
dilepton spectra at heavy-ion collider experiments at RHIC and LHC. This includes 
invariant-mass spectra at full RHIC energy with transverse-momentum dependencies 
and azimuthal asymmetries, as well as a systematic investigation of the excitation 
function down to fixed-target energies, thus making contact to previous precision 
measurements at the SPS. Furthermore, predictions for the energy frontier at the 
LHC are presented in both dielectron and dimuon channels.      
\end{abstract}

\maketitle

%%%%%%%%%%%%%%%%%%%%%%%%%%%%%%%%%%%%%%%%%%%%
%% MAINMATTER
%%%%%%%%%%%%%%%%%%%%%%%%%%%%%%%%%%%%%%%%%%%%

%%%%%%%%%%%%%%%%%%%%%%%%%%%%%%%
\section{Introduction}
\label{sec_intro}
%%%%%%%%%%%%%%%%%%%%%%%%%%%%%%%%
The exploration of matter at extremes of temperature ($T$) and baryon density 
($\varrho_B$) is at the forefront of research in contemporary nuclear physics, 
with intimate connections to high-energy, condensed-matter and even atomic 
physics~\cite{Shuryak:2008eq}. 
Theoretical efforts over the last few decades are suggesting an 
extraordinary richness of the phase diagram of strongly interacting matter, 
which should ultimately emerge from the underlying theory of Quantum 
Chromodynamics (QCD) as part of the Standard Model. However, several basic 
questions, both qualitative and quantitative, such as the possible existence 
of first order transitions and their location as function of baryon-chemical 
potential ($\mu_B$) and temperature, remain open to date~\cite{Friman:2011zz}. 
A close interplay of experiment and theory is needed to create a robust 
knowledge about the QCD phase structure. On the one 
hand, naturally occurring  matter at temperatures close to or beyond the 
expected pseudo-critical one, 
$T_{\rm pc}\simeq160$\,MeV~\cite{Borsanyi:2010bp,Bazavov:2011nk}, may last 
have existed $\sim$14 billion years ago, during the first tens of microseconds 
of the Universe. On the other hand, at small temperatures, matter with 
baryon densities close to or beyond the critical one for the transition 
into quark matter may prevail in the interior of compact stars today, but 
its verification and exploration from observational data is 
challenging~\cite{Weber:2004kj}. It is quite fascinating that tiny man-made 
samples of hot QCD matter can nowadays be created and studied in the 
laboratory using ultrarelativistic heavy-ion collisions (URHICs).
Significant progress has been made in understanding the properties of this 
medium through analyses of experiments conducted at the CERN's Super-Proton
Synchrotron (SPS), BNL's Relativistic Heavy-Ion Collider (RHIC) and CERN's 
Large Hadron Collider (LHC) (see, e.g., the recent Quark Matter conference 
proceedings~\cite{qm11,qm12}). For example, systematic investigations of the 
produced hadron spectra have revealed a hydrodynamic behavior of the bulk 
matter in the region of low transverse momenta 
($q_t\lsim$~2-3\,GeV) and a strong absorption of hadrons with high transverse 
momentum ($q_t\gsim$6\,GeV). Even hadrons containing a heavy quark (charm or 
bottom) exhibit substantial energy loss and collectivity due to their coupling 
to the expanding fireball. While the total charm and bottom yields are 
essentially conserved, the production of heavy quark-antiquark bound states 
(charmonia and bottomonia) is largely suppressed. The relation of the above 
hadronic observables to spectral properties of the medium is, however, rather 
indirect. Low-mass dileptons, on the other hand, are radiated from the interior 
of the medium throughout the fireball's lifetime, as their mean-free path is much 
larger than the size of the fireball. Thus, their invariant-mass spectra directly 
measure the in-medium vector spectral function, albeit in a superposition of the 
varying temperature in the fireball's expansion.       

The dilepton program at the SPS has produced remarkable results. The
CERES/NA45 dielectron data in Pb-Au collisions, and particularly the NA60 
dimuon spectra in In-In collisions, have shown that the $\rho$-meson undergoes
a strong broadening, even complete melting, of its resonance structure, with
quantitative sensitivity to its spectral shape, see 
Refs.~\cite{Tserruya:2009zt,Specht:2010xu,Rapp:2009yu} for recent reviews.
The QCD medium at SPS energies is characterized by a significant net-baryon
content with chemical potentials of $\mu_B\simeq 250$\,MeV at chemical freezeout,
$T_{\rm ch}\simeq160$\,MeV~\cite{BraunMunzinger:2011ze}, and further increasing 
as the system cools down~\cite{Rapp:2002fc}. Baryons have been identified as a 
dominant contributor to the medium modifications of the $\rho$'s spectral 
function~\cite{Rapp:2009yu}. The question arises how these develop when moving 
toward the net 
baryon-free regime in the QCD phase diagram, $\mu_B\ll T$. Theoretical 
expectations based on the hadronic many-body approach~\cite{Rapp:2000pe} suggest 
comparable medium effects in this regime, since the relevant quantity is the 
{\em sum} of baryon and antibaryon densities, and this turns out to be similar 
at SPS and RHIC/LHC~\cite{Rapp:2002fc}, at least close to $T_{\rm pc}$. Since 
$T_{\rm ch}\simeq T_{\rm pc}$ at collider energies, the total baryon density 
at RHIC and LHC in the subsequent hadronic evolution of the fireball will remain
similar. We also note that the $\mu_B\simeq 0$\,MeV regime is amenable to 
numerical lattice QCD calculations, both for the equation of state of the medium 
evolution, and in particular for the microscopic dilepton production rate,
at least in the QGP phase for now~\cite{Ding:2010ga,Brandt:2012jc}.
Furthermore, since the phase transition at $\mu_B\simeq 0$\,MeV presumably is
a continuous crossover~\cite{Aoki:2006we}, a realistic dilepton rate should vary 
smoothly when 
changing the temperature through $T_{\rm pc}$. Thus, after the successful 
fixed-target dilepton program at the CERN-SPS, the efforts and attention are 
now shifting to collider energies around experiments at RHIC and LHC.  

In the present article we will focus on the theory and phenomenology of 
dilepton production at collider energies (for a recent overview including 
an assessment of SPS data, see, e.g., Ref.~\cite{Rapp:2012zq}). The presented 
material is partly of 
review nature, but also contains thus far unpublished results, e.g., updates in 
the use of nonperturbative QGP dilepton rates and equation of state, and detailed 
predictions for invariant-mass and transverse-momentum spectra for ongoing and 
upcoming experiments at RHIC and LHC, including an excitation function of the 
beam energy scan program at RHIC.

This article is organized as follows. In Sec.~\ref{sec_rates} we briefly review 
the calculation of the thermal dilepton emission rates from hadronic matter and 
the quark-gluon plasma (QGP). We elaborate on how recent lattice-QCD results at 
vanishing three-momentum ($q$=0) may be extended to finite $q$ to enable their 
application to URHICs. In Sec.~\ref{sec_spectra} we discuss in some detail the 
calculations of dilepton spectra suitable for comparison with experiment; this 
involves a brief discussion of the medium evolution in URHICs (including an 
update of the equation of state) in Sec.~\ref{ssec_evo}, and of 
non-thermal sources (primordial production and final-state decays) in 
Sec.~\ref{ssec_non-thermal}. It will be followed by analyses of mass and 
momentum spectra, as well as elliptic flow at full RHIC energy in 
Sec.~\ref{ssec_rhic200}, and of an excitation function as obtained from the 
RHIC beam energy scan in Sec.~\ref{ssec_rhic-bes};
predictions for dielectron and dimuon spectra at current (2.76~ATeV) and 
future (5.5~ATeV) LHC energies are presented in Sec.~\ref{ssec_lhc}. We end 
with a summary and outlook in Sec.~\ref{sec_sum}. 

%%%%%%%%%%%%%%%%%%%%%%%%%%%%%%%%%%%%%%%%%%%%%%%%%%%%%%
\section{Thermal Dilepton Rates in QCD Matter}
\label{sec_rates}
%%%%%%%%%%%%%%%%%%%%%%%%%%%%%%%%%%%%%%%%%%%%%%%%%%%%%%
The basic quantity for connecting calculations of the electromagnetic (EM) 
spectral function in QCD matter to measurements of dileptons in heavy-ion 
collisions is their thermal emission rate; per unit phase space it can be 
written as
\begin{equation}
\frac{dN_{ll}}{d^4xd^4q} = -\frac{\alpha_{\rm EM}^2 L(M)}{\pi^3 M^2} \ 
f^B(q_0;T) \ {\rm Im}\Pi_{\rm EM}(M,q;\mu_B,T) \ ,
\label{rate}
\end{equation}
where $L(M)$ is a lepton phase-space factor (=1 for vanishing lepton mass),
$f^B$ denotes the thermal Bose distribution, and $q_0=\sqrt{M^2+q^2}$  is
the energy of the lepton pair (or virtual photon) in terms of
its invariant mass and 3-momentum. As mentioned above, this observable is
unique in its direct access to an in-medium spectral function of the formed
system, namely in the vector (or EM) channel,
${\rm Im}\Pi_{\rm EM}\equiv \frac{1}{3} g_{\mu\nu} {\rm Im}\Pi_{\rm EM}^{\mu\nu}$. 
It is defined via the correlation function of the EM current, 
$j^\mu_{\rm EM}$, as transported by the electric-charge carriers 
in the system. 
In quark basis, the EM current is given by the charge-weighted sum over flavor, 
\begin{equation}
j^\mu_{\rm EM} = \sum\limits_{q=u,d,s} e_q \ \bar{q} \gamma^\mu q \ ,
\label{j-quark}
\end{equation}
while in hadronic basis it is in good approximation given by the 
vector-meson fields, 
\begin{equation}
 j^\mu_{\rm EM}= \sum\limits_{V=\rho,\omega,\phi}\frac{m_V^2}{g_V} V^\mu \, , 
\label{vdm}
\end{equation}
known as vector-dominance model (VDM).
Since the significance of thermal dilepton radiation is limited to masses 
below the $J/\psi$ mass, $M\lsim3$\,GeV, we will focus on the light- and
strange-quark sector in this article. 

\begin{figure}[!t]
\begin{minipage}{1.0\linewidth}
\begin{center}
  \includegraphics[width=0.8\textwidth,angle=0]{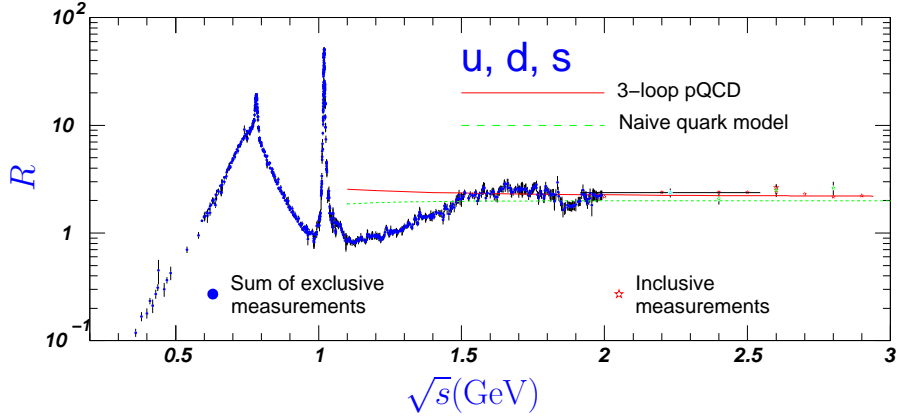}
\end{center}
\vspace{-1cm}
\end{minipage}
\caption{Compilation of experimental data for the ratio, $R$, of cross sections
for $e^+e^-\to hadrons$ over $e^+e^-\to\mu^+\mu^-$, as a function of
invariant mass $\sqrt{s}=M$. Figure taken from Ref.~\cite{Nakamura:2010zzi}.}
\label{fig_Rhad}
\end{figure}
In the vacuum, the EM spectral function is well known from the $e^+e^-$
annihilation cross section into hadrons, usually quoted relative to the
annihilation into dimuons as the ratio
$R=-\frac{12\pi}{s} {\rm Im}\Pi_{\rm EM}$, cf.~Fig.~\ref{fig_Rhad}.
It illustrates that the nonperturbative hadronic description in terms of 
VDM works well in the low-mass region (LMR), 
$M\lsim1$\,GeV, while the perturbative partonic description appears to apply for 
$M\gsim1.5$\,GeV.  Thus, in URHICs, dilepton spectra in the LMR are ideally 
suited to study the properties of vector mesons in the medium. A central 
question is if and how these medium modifications can signal (the approach to) 
deconfinement and the restoration of the dynamical breaking of chiral symmetry 
(DBCS). After all, confinement and DBCS govern the properties of hadrons in 
vacuum. At masses $M\gsim1.5$\,GeV, the perturbative nature 
of the EM spectral function suggests that in-medium modifications are suppressed, 
coming in as corrections in powers of $T/M$ and $\alpha_s$.   
In this case, invariant-mass spectra of thermal radiation become an excellent 
measure for the prevalent temperatures of the produced system, free from blue 
shifts due to the medium expansion which strongly affect $p_t$ spectra.

%%%%%%%%%%%%%%%%%%%%%%%%%%%%%%%%%%%%%%%%%%%%%%%%%%%%%%
\subsection{Hadronic Matter}
\label{ssec_hm}
%%%%%%%%%%%%%%%%%%%%%%%%%%%%%%%%%%%%%%%%%%%%%%%%%%%%%%
Over the last two decades, broad efforts have been undertaken to evaluate the 
medium modifications of the $\rho$-meson. The latter dominates in the EM spectral 
function over the $\omega$ by about a factor of 10 (the $\phi$ appears to be 
rather protected from hadronic medium effects, presumably due to the OZI rule, 
at least for its coupling to baryons). Recent overviews of these efforts can be 
found, e.g., in Refs.~\cite{Rapp:2009yu,Leupold:2009kz,Oset:2012ap}. Most 
approaches utilize effective hadronic (chiral) Lagrangians and apply them in 
diagrammatic many-body theory to compute thermal (or density) loop corrections.
The generic outcome is that of a substantial broadening of the $\rho$'s 
spectral shape, with little mass shift (in a heat bath, chiral symmetry protects
the $\rho$ from mass shifts at order ${\cal O}(T^2)$~\cite{Dey:1990ba}). The 
magnitude of the $\rho$'s in-medium width (and/or its precise spectral shape) 
varies in different calculations, but the discrepancies can be mostly traced 
back to the differing contributions accounted for in the Lagrangian (e.g., 
the set of baryon and/or meson resonance excitations, or medium effects in the 
$\rho$'s pion cloud).  
Similar findings arise when utilizing empirically extracted on-shell 
$\rho$-meson scattering amplitudes off hadrons in linear-density 
approximation~\cite{Eletsky:2001bb}. Since these calculations are restricted
to resonances above the nominal $\rho N$ (or $\rho\pi$) threshold, 
quantitative differences to many-body (field-theoretic) approaches may 
arise. In particular, the latter account for subthreshold excitations, 
e.g., $\rho+N \to N^*(1520)$, which induce additional broadening and
associated enhancement of the low-mass part in the $\rho$ spectral
function (also causing marked deviations from a Breit-Wigner shape).  
Appreciable mass shifts are typically found in mean-field approximations 
(due to large in-medium scalar fields), or in calculations where the bare 
parameters of the underlying Lagrangian are allowed to be 
temperature dependent~\cite{Harada:2003jx}. 

\begin{figure}[!t]
\begin{minipage}{0.5\linewidth}
\vspace{0.1cm}
 \includegraphics[width=0.97\textwidth]{drdm2-Inx160.eps}
\end{minipage}
\begin{minipage}{0.5\linewidth}
%\vspace{-0.4cm}
  \includegraphics[width=0.91\textwidth,angle=-90]{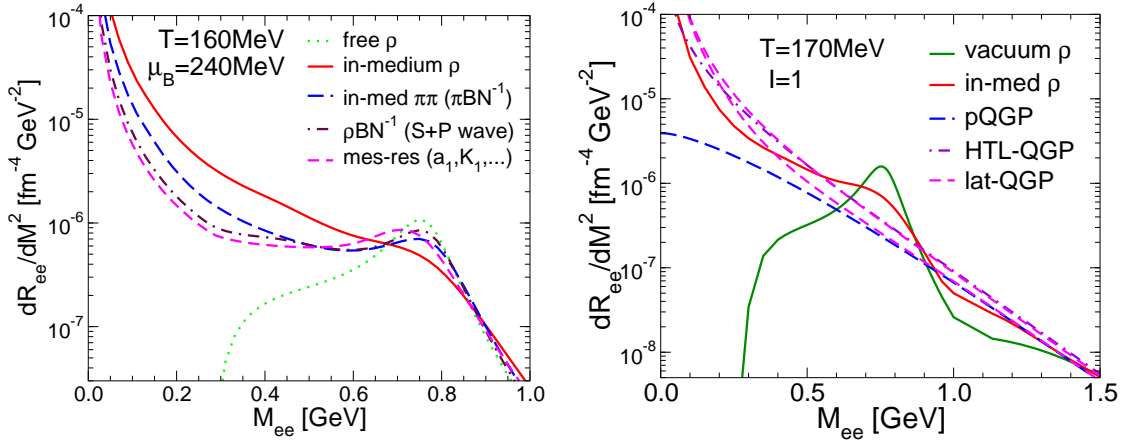}
\end{minipage}
\caption{Dilepton rates from hot QCD matter in the isovector ($\rho$) channel. 
Left panel: effective hadronic Lagrangian plus many-body approach for the 
in-medium $\rho$ spectral function (solid line) at a temperature and chemical
potential characteristic for chemical freezeout at full SPS energy; the effects
of in-medium pion-cloud (long-dashed line), baryon resonances (dash-dotted line)
and meson resonances (short-dashed line) are shown separately along
with the rate based on the vacuum spectral function (dotted line). 
Right panel: comparison of free and in-medium hadronic and partonic calculations 
at temperature $T=170$~MeV and small baryon chemical potential
characteristic for RHIC and LHC conditions; the free and in-medium hadronic 
rates are based on Refs.~\cite{Urban:1998eg,Rapp:1999us}; the "lat-QGP" rates
(2 short-dashed lines) are based on fits to the $q$=0 lQCD rate with
extensions to finite 3-momentum utilizing perturbative photon rates
(see Sec.~\ref{ssec_qgp} for details).}
\label{fig_rates}
\end{figure}
An example for dilepton rates following from a $\rho$ spectral function 
calculated in hot and dense hadronic matter at SPS energies is shown in the 
left panel of Fig.~\ref{fig_rates}. The EM spectral function follows from 
the $\rho$-meson using VDM, Eq.~(\ref{vdm}), although corrections to VDM 
are necessary for quantitative descriptions of
the EM couplings in the baryon sector~\cite{Friman:1997tc,Rapp:1997fs}.
When extrapolated to temperatures around $T_{\rm pc}$, the resonance peak  
has essentially vanished leading to a structureless emission rate with a large
enhancement in the mass region below the free $\rho$ mass.  
%marked deviations from a Breit-Wigner shape are caused 
The decomposition of the rate into in-medium self-energy contributions 
illustrates the important role of the pion cloud modifications, and of 
multiple low-energy excitations below the free $\rho$ mass, e.g.,  
resonance-hole $BN^{-1}$, i.e., $\rho+N\to B$ for off-shell $\rho$-mesons. 
The hadronic medium effects are slightly reduced at collider energies (right 
panel of Fig.~\ref{fig_rates}), where a faint resonance structure appears to
survive at around $T_{\rm pc}$ (it is significantly more suppressed at
$T$=180\,MeV). 
A recent calculation in a similar framework, combing thermal field 
theory with effective hadron Lagrangians~\cite{Ghosh:2011gs}, and including both
finite-temperature and -density contributions to the $\rho$ selfenergy
through baryon and meson resonances, shows fair agreement with the results 
shown in the left panel of Fig.~\ref{fig_rates}.

%%%%%%%%%%%%%%%%%%%%%%%%%%%%%%%%%%%%%%%%%%%%%%%%%%%%%%
\subsection{Quark-Gluon Plasma}
\label{ssec_qgp}
%%%%%%%%%%%%%%%%%%%%%%%%%%%%%%%%%%%%%%%%%%%%%%%%%%%%%%
In a perturbative QGP (pQGP), the leading-order (LO) mechanism of 
dilepton production 
is EM quark-antiquark annihilation as following from a free quark current in 
Eq.~(\ref{j-quark}). The corresponding EM spectral function is essentially given 
by the ``naive quark model" curve in Fig.~\ref{fig_Rhad}, 
extended all the way down to vanishing mass, 
\begin{equation}
{\rm Im}~\Pi_{\rm EM}^{\rm pQGP} = - \frac{C_{\rm EM} N_c}{12\pi} M^2  
\left(1+\frac{2T}{q} \ln[\frac{1+x_+}{1+x_-}] \right) \equiv 
- \frac{C_{\rm EM} N_c}{12\pi} M^2 \hat{f}_2(q_0,q;T) ,
\end{equation}
where $C_{\rm EM}\equiv \sum_{q=u,d,s} e_q^2$ 
(an additional phase-space factor occurs for finite current quark masses) and 
$x_\pm=\exp[-(q_0\pm q)/2T]$. Finite-temperature corrections are induced 
by a quantum-statistical Pauli-blocking factor (written for $\mu_q$=0) which 
produces a nontrivial 3-momentum 
dependence~\cite{Cleymans:1986na}; for $q$=0 it simplifies 
to $\hat{f}_2(q_0,q=0;T)=[1-2f^F(q_0/2)]$, 
where $f^F$ is the thermal Fermi distribution. The pertinent 3-momentum
integrated dilepton rate is structureless, cf.~long-dashed curve in 
Fig.~\ref{fig_rates} right. It's finite value at $M=0$ implies
that no real photons can be produced from this mechanism.   

A consistent implementation of $\alpha_s$ corrections in a thermal QGP at
vanishing quark chemical potential has been achieved by resumming the 
hard-thermal-loop (HTL) action~\cite{Braaten:1990wp}. Quarks and gluons 
acquire thermal masses $m_{q,g}^{th}\sim gT$, but Bremsstrahlungs-type 
contributions lead to a marked enhancement of the rate over the LO pQCD 
results, cf.~the dash-dotted line in the right panel of Fig.~\ref{fig_rates}.

\begin{figure}[!t]
\begin{minipage}{0.5\linewidth}
\vspace{0.3cm}
  \includegraphics[width=0.89\textwidth]{Pii180.eps}
\end{minipage}
\begin{minipage}{0.5\linewidth}
\vspace{-0.4cm}
  \includegraphics[width=0.86\textwidth,angle=-90]{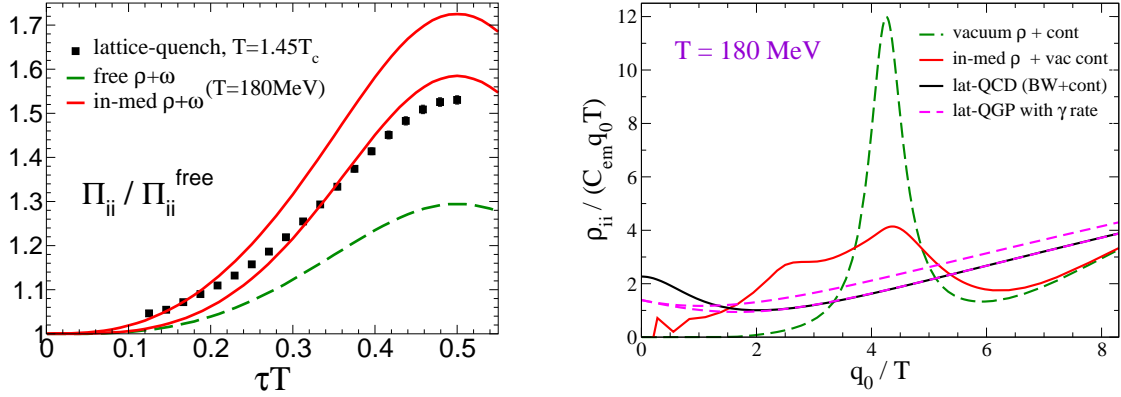}
\end{minipage}
\caption{Left panel: Euclidean correlators of the EM current as computed in 
quenched thermal lQCD (data points)~\cite{Ding:2010ga}, compared to results 
from integrating hadronic spectral functions using Eq.~(\ref{Pi-tau}) without
(dashed green line) and with in-medium effects (red lines, with free and 
in-medium continuum threshold)~\cite{Rapp:2002pn}.  
Right panel: Vector-isovector spectral functions at $q$=0 corresponding
to the euclidean correlators in the left panel: in vacuum (green dashed line), in 
hadronic matter calculated from many-body theory at $T$=180\,MeV~\cite{Rapp:2000pe} 
(red solid line) and in a gluon plasma at 1.4\,$T_c$ extracted from thermal 
lattice-QCD (black solid line)~\cite{Ding:2010ga}; the 3-momentum extended lQCD 
rates according to Eq.~(\ref{PiT-LAT}) are shown for $K$=2 (short-dashed lines, 
with (lower) and without (upper) formfactor correction).}
\label{fig_sf}
\end{figure}
Recent progress in calculating dilepton rates nonperturbatively using thermal
lattice QCD (lQCD) has been reported in 
Refs.~\cite{Ding:2010ga,Brandt:2012jc,Ding:2013qw}. The basic quantity 
computed in these simulations is the euclidean-time correlation function 
which is related to the spectral function, $\rho_V\equiv -2{\rm Im}\Pi_i^i$, 
via \begin{equation}
\Pi_V(\tau,q;T)=
\int\limits_0^\infty \frac{dq_0}{2\pi} \ \rho_V(q_0,q;T) \
\frac{\cosh[q_0(\tau-1/2T)]}{\sinh[q_0/2T]} \ .
\label{Pi-tau}
\end{equation}
Results for $\Pi_V$ obtained in quenched QCD for $T$=1.45$T_c$ at vanishing
$q$ (in which case $M$=$q_0$) are shown by the data points in the left panel 
of Fig.~\ref{fig_sf}, normalized to the free (non-interacting) pQGP limit. 
At small $\tau$, corresponding to large energies in 
the spectral function, this ratio tends to one as expected for the 
perturbative limit. For larger $\tau$, a 
significant enhancement develops which is associated with a corresponding
enhancement in the low-energy (or low-mass) regime of the spectral function
(and thus dilepton rate). This enhancement may be quantified by making an 
ansatz for the spectral function in terms of a low-energy Breit-Wigner part 
plus a perturbative continuum~\cite{Ding:2010ga}, 
\begin{equation}
\rho_V^{ii}(q_0) = S_{\rm BW} \frac{q_0 \Gamma/2}{q_0^2+\Gamma^2/4} 
                +\frac{C_{\rm EM}N_c}{2\pi} (1+\kappa) q_0^2 \tanh(q_0/4T) \ 
\label{rhoV_lat}
\end{equation}   
(note that $\tanh(q_0/4T)=1-2f^F(q_0/2)$).
The strength ($S_{\rm BW}$) and width ($\Gamma$) of the Breit-Wigner,
as well as a perturbative $\alpha_s$ correction ($\kappa$), are then fit
to the euclidean correlator. The large-$\tau$ enhancement in the correlator 
generates an appreciable low-energy enhancement in the spectral function, 
cf.~right panel of Fig.~\ref{fig_sf}. The zero-energy limit of the spectral 
function defines a transport coefficient, the electric conductivity, 
$\sigma_{\rm EM}=\frac{1}{6}\lim_{q_0\to0}(\rho_V^{ii}/q_0)$. 
Similar to the viscosity or heavy-quark diffusion coefficient, a small value 
for $\sigma_{\rm EM}$, implied by a large value for $\Gamma$, indicates a 
strong coupling of the medium; e.g., in pQCD, 
$\sigma_{\rm EM}\propto T/\alpha_s^2$~\cite{Moore:2006qn}.  
The results for the dilepton rate (or spectral function) at a smaller temperature 
of 1.1\,$T_c$ are found to be similar to the ones at 
1.45\,$T_c$~\cite{Ding:2013qw}, suggesting a weak temperature dependence in this 
regime. Note, however, that the phase transition in quenched QCD is of first 
order, i.e., a stronger variation is expected when going across $T_c$. 
Recent results for two-flavor QCD~\cite{Brandt:2012jc} also indicate rather 
structureless spectral functions similar to the quenched results.
Ultimately, at sufficiently small temperatures, the lattice computations 
should recover a $\rho$-meson resonance peak; it will be interesting to see
at which temperatures this occurs. 

For practical applications, a finite 3-momentum dependence of the lQCD 
dilepton rate is needed, which is currently not available from the 
simulations. We here propose a ``minimal" construction 
which is based on a matching to the 3-momentum dependence obtained from 
the LO pQCD photon rate~\cite{Kapusta:1991qp}. The latter reads
\begin{eqnarray}
q_0 \frac{dR_\gamma}{d^3q} &=& 
-\frac{\alpha_{\rm EM}}{\pi^2} {\rm Im} \Pi_T(M=0,q) f^B(q_0,T)
\nonumber\\
&=& \frac{C_{\rm EM}\alpha\alpha_S}{2\pi^2} T^2 f^B(q_0,T) 
%{\rm e}^{-q_0/T}
       \ln\left(1+\frac{2.912}{4\pi\alpha_s} \frac{q_0}{T} \right) \  . 
\label{rate_pqcd}
\end{eqnarray}
The idea is now to adopt the transverse part of the EM spectral function as
given by Eq.~(\ref{rate_pqcd}) for the 3-momentum dependence of the spectral
function in Eq.~(\ref{rhoV_lat}) by replacing the Breit-Wigner part with it, 
i.e.,
\begin{eqnarray}
 -{\rm Im}~\Pi_T &=& \frac{C_{\rm EM} N_c}{12\pi} M^2 
\left(\hat{f}_2(q_0,q;T) + 2\pi \alpha_s \frac{T^2}{M^2} K\,F(M^2) 
\ln\left(1+\frac{2.912}{4\pi\alpha_s} \frac{q_0}{T} \right) \right) 
\nonumber\\ 
&\equiv& \frac{C_{\rm EM} N_c}{12\pi} M^2 
\left(\hat{f}_2(q_0,q;T)+Q_{\rm LAT}^T(M,q)\right) \ . 
\label{PiT-LAT}
\end{eqnarray}
Here we have introduced a $K$ factor into $Q_{\rm LAT}^T$, which serves 
two purposes: (i) With $K$=2 it rather accurately accounts for the enhancement 
of the complete LO photon rate calculation~\cite{Arnold:2001ms} over the rate
in Eq.~(\ref{rate_pqcd}); (ii) It better reproduces the low-energy regime of 
the lQCD spectral function; for example, for $K$=2 the electric conductivity 
following from Eq.~(\ref{PiT-LAT}) is $\sigma_{\rm EM}/T \simeq 0.23 C_{\rm EM}$,  
not far from the lQCD estimate with the fit ansatz (\ref{rhoV_lat}), 
$\sigma_{\rm EM}/T \simeq (0.37\pm0.01) C_{\rm EM}$ (also
compatible with Ref.~\cite{Aarts:2007wj}; the systematic uncertainty in the
lattice result, due to variations in the ansatz, is significantly larger).
The resulting spectral function (upper dashed line in Fig.~\ref{fig_sf} right)
somewhat overestimates the lQCD result at high energies, where the latter 
coincides with the annihilation term. This can be improved by an additional 
formfactor, $F(M^2)=\Lambda^2/(\Lambda^2+M^2)$, resulting in the lower
dashed line in the right panel of Fig.~\ref{fig_sf} (using $\Lambda=2T$). 

Finally, care has to be taken to include a finite longitudinal part which 
develops in the timelike regime. Here we employ a dependence that follows, 
e.g., from standard constructions of gauge-invariant $S$-wave $\rho$-baryon 
interactions, yielding $\Pi_L = (M^2/q_0^2) \Pi_T$~\cite{Rapp:1999ej}. Thus, 
we finally have
\begin{equation}
Q_{LAT}^{\rm tot} = \frac{1}{3} \left(2Q_{LAT}^{T} + Q_{LAT}^{L}\right)  
        = \frac{1}{3} Q_{LAT}^{T} \left( 2 + \frac{M^2}{q_0^2} \right) \ .
\end{equation} 

The lQCD results for the isovector spectral function are compared to 
hadronic calculations in the right panel of Fig.~\ref{fig_sf}. 
Close to the phase transition temperature, the ``melting" of
the in-medium $\rho$ spectral function suggests a smooth transition from
its prominent resonance peak in vacuum to the rather structureless
shape extracted from lQCD, signaling a transition from hadronic to partonic 
degrees of freedom. It would
clearly be of interest to extract the conductivity from the hadronic
calculations, which currently is not well resolved from the $q$=0, $q_0\to0$
limit of the spectral function. The mutual approach of the nonperturbative
hadronic and lQCD spectral functions is also exhibited in the 3-momentum
integrated dilepton rate shown in the right panel of Fig.~\ref{fig_rates},
especially when compared to the different shapes of the LO pQCD 
and vacuum hadronic rates. Arguably, the in-medium hadronic rate still shows
an indication of a broad resonance. A smooth matching of the rates
from above and below $T_{\rm pc}$ might therefore require some additional 
medium effects in the hot and dense hadronic medium, and/or the emergence 
of resonance correlations in the $q\bar q$ correlator in the QGP. 
Unless otherwise noted, the thermal emission rates used in the calculations
of dilepton spectra discussed below will be based on the in-medium
hadronic rates of Ref.~\cite{Rapp:1999us} and the lQCD-inspired QGP 
rates~\cite{Ding:2010ga}, extended to finite 3-momentum as constructed 
above (with $K$=2 and formfactor). 

%%%%%%%%%%%%%%%%%%%%%%%%%%%%%%%%%%%%%%%%%%%%%%%%%%%%%%
\section{Dilepton Spectra at RHIC and LHC}
\label{sec_spectra}
%%%%%%%%%%%%%%%%%%%%%%%%%%%%%%%%%%%%%%%%%%%%%%%%%%%%%%
The calculation of dilepton mass and transverse-momentum ($q_t$) spectra, 
suitable for comparison to data in heavy-ion collisions, requires an
integration of the thermal rates of hadronic matter and QGP over a realistic 
space-time evolution of the AA reaction, 
\begin{equation}
\frac{dN_{ll}}{dM} = \int d^4x \ \frac{Md^3q}{q_0} \ 
\frac{dN_{ll}}{d^4xd^4q} \ .
\end{equation}
In addition to the thermal yield, non-thermal sources 
have to be considered, e.g., primordial Drell-Yan annihilation and 
electromagnetic final-state decays of long-lived hadrons. We will briefly 
discuss space-time evolutions in Sec.~\ref{ssec_evo} and non-thermal
sources in Sec.~\ref{ssec_non-thermal},  
before proceeding to a more detailed discussion of thermal spectra and 
comparisons to data, as available, in Secs.~\ref{ssec_rhic200}, 
\ref{ssec_rhic-bes} and \ref{ssec_lhc} for full RHIC energy, the beam-energy
scan and LHC, respectively.

%%%%%%%%%%%%%%%%%%%%%%%%%%%%%%%%%%%%%%%%%%%%%%%%%%%%%
\subsection{Medium Expansion}
\label{ssec_evo}
%%%%%%%%%%%%%%%%%%%%%%%%%%%%%%%%%%%%%%%%%%%%%%%%%%%%%%
The natural framework to carry out the space-time integral over the 
dilepton rate in URHICs is relativistic hydrodynamics. The application
of this approach to AA collisions at RHIC and LHC works well to 
describe bulk hadron observables (e.g., $p_t$ spectra and elliptic flow) 
up to momenta of $p_t\simeq$~2-3\,GeV, which typically comprises more than
90\% of the total yields. Some uncertainties remain, e.g., as to the precise
initial conditions at thermalization, viscous corrections, or the treatment 
of the late stages where the medium becomes dilute and the hadrons decouple
(see, e.g., Ref.~\cite{Heinz:2013th} for a recent review). 
Another key ingredient is the equation of state (EoS) of the medium, 
$\epsilon(P)$, which drives its collective expansion.   
\begin{figure}[!t]
\begin{minipage}{0.55\linewidth}
\vspace{0.5cm}
\includegraphics[width=1.05\textwidth,angle=-0]{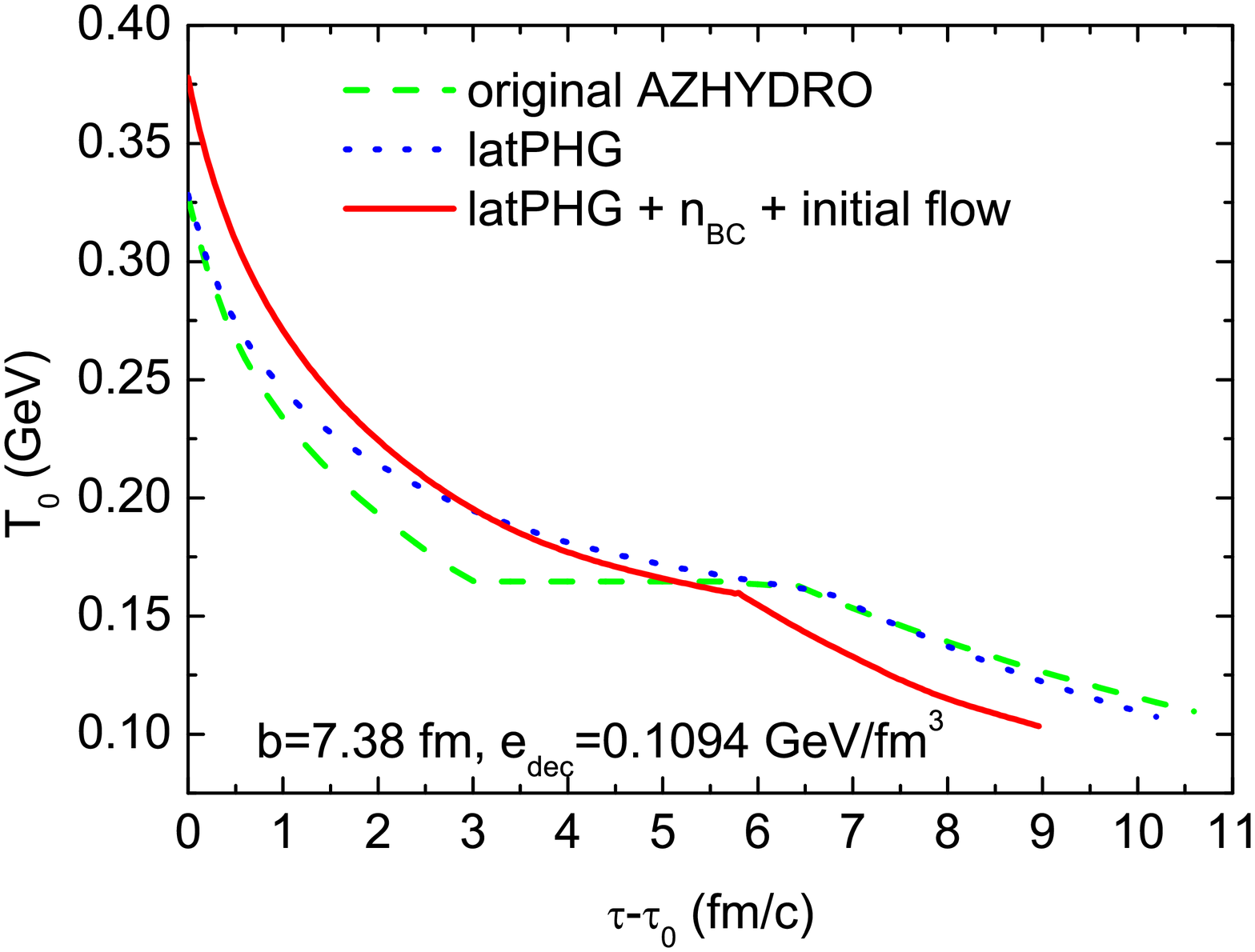}
\end{minipage}
\hspace{-0.4cm}
\begin{minipage}{0.5\linewidth}
\vspace{0.3cm}
\includegraphics[width=0.82\textwidth,angle=-90]{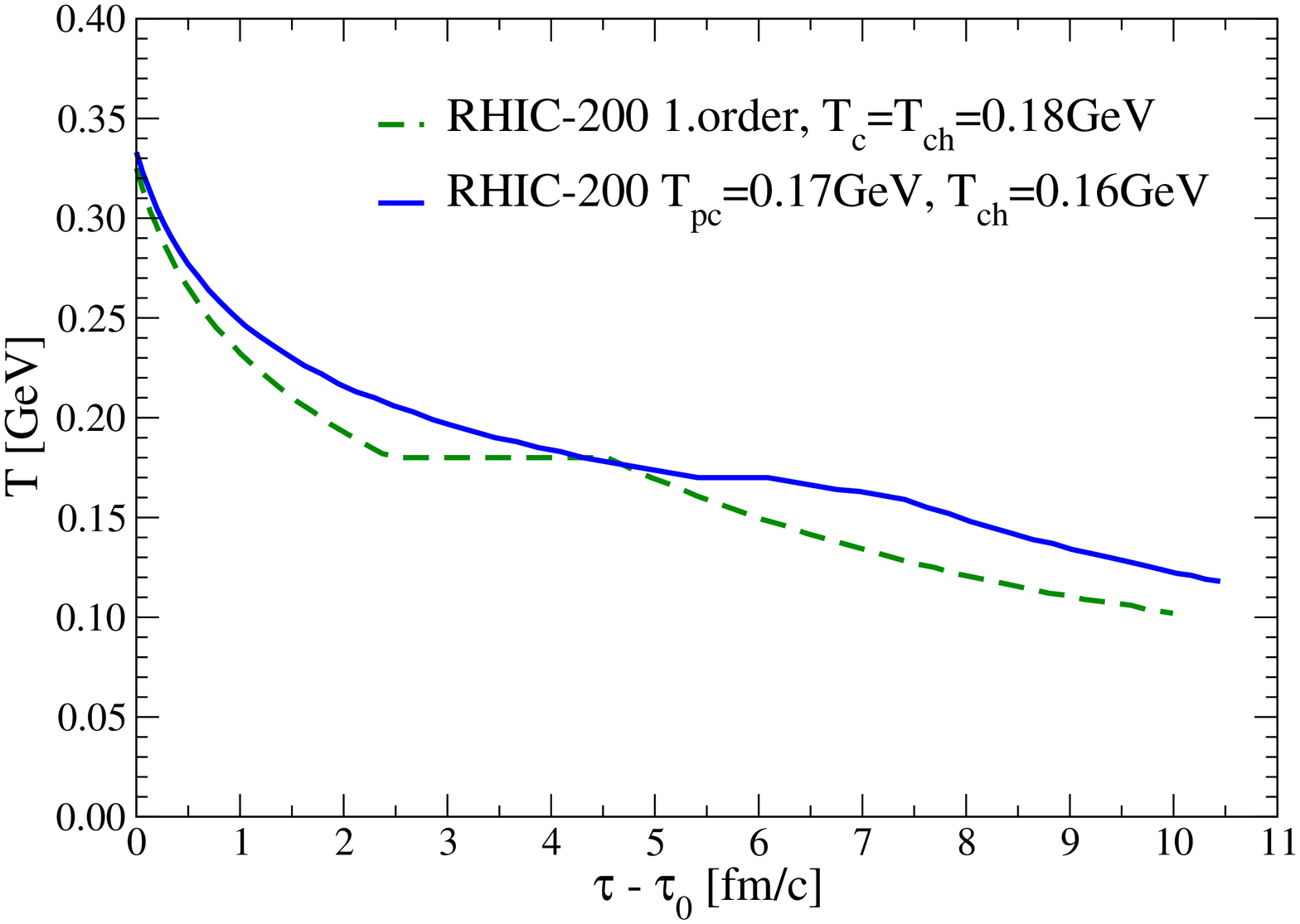}
\end{minipage}
\caption{Time evolution of fireball temperature in semicentral 
Au-Au($\sqrt{s}$=0.2\,GeV) collisions at RHIC within (the central cell of) 
ideal hydrodynamics (left)~\cite{He:2011zx} and an expanding fireball model 
(right). The dashed green and dotted blue line in the left panel are to be compared
the dashed green and solid blue line in the right panel, respectively.}
\label{fig_temp}
\end{figure}
The left panel of Fig.~\ref{fig_temp} illustrates the effects of updating a 
previously employed bag-model EoS (a quasiparticle QGP connected to a hadron 
resonance gas via a first-order phase transition)~\cite{Kolb:2003dz} by a recent 
parametrization of a nonperturbative QGP EoS from lQCD 
data~\cite{Borsanyi:2010cj,Cheng:2009zi} (continuously 
matched to a hadron-resonance gas at $T_{\rm pc}$=170\,MeV)~\cite{He:2011zx}: 
within a 2+1-D ideal hydro calculation the most notable change is a 
significant increase of the temperature (at fixed entropy density)
in the regime just above the transition temperature (up to ca.~30~MeV
at the formerly onset of the first-order transition). Together with the
fact that the hadronic portion of the formerly mixed phase is now entirely
associated with the QGP, this will lead to an increase (decrease) of the 
QGP (hadronic) contribution to EM radiation relative to the first-order
scenario. In addition, the
harder lattice EoS induces a stronger expansion leading to a slightly
faster cooling and thus reduction in the lifetime by about 5\%. This
effect becomes more pronounced when modifying the initial conditions
of the hydrodynamic evolution, e.g., by introducing a more 
compact spatial profile (creating larger gradients) and/or initial 
transverse flow (associated with interactions prior to the thermalization 
time, $\tau_0$)~\cite{He:2011zx}, cf.~the solid line in Fig.~\ref{fig_temp}
left. The resulting more violent expansion plays an important 
role in understanding the HBT radii of the 
system~\cite{Pratt:2008qv}. The relevance for EM radiation pertains to 
reducing the fireball lifetime by up to $\sim$20\%.

More simplistic parametrizations of the space-time evolution of AA
collisions have been attempted with longitudinally and transversely
expanding fireballs. With an appropriate choice of the transverse acceleration
(in all applications below it is taken as $a_t$=0.12/fm at the surface),
an approximate reproduction of the basic features (timescales and radial
flow) of hydrodynamic evolutions can be achieved, see
right panel of Fig.~\ref{fig_temp}. Most of the results shown 
in the remainder of this article are based on such simplified fireball 
parametrizations, utilizing the EoS of Ref.~\cite{He:2011zx}
where a parametrization of lQCD results is matched with a hadron resonance
gas at $T_{\rm pc}$=170\,MeV and subsequent chemical freezeout at 
$T_{\rm ch}$=160\,MeV (see also Ref.~\cite{Renk:2002md}). We note that the 
use of this EoS, together with the lQCD-based QGP emission rates, constitutes 
an update of our earlier 
calculations~\cite{vanHees:2007th} where a quasiparticle bag-model 
EoS was employed in connection with HTL rates in the QGP. We have checked that 
the previous level of agreement with the acceptance-corrected NA60 spectra is 
maintained, which is essentially due to the duality of the QGP and hadronic 
rates around $T_{\rm pc}$ (a more detailed account in the context of the SPS
dilepton data will be given elsewhere~\cite{vanHees:2013}). For our discussion
of collider energies below, the initialization (or thermalization times) are chosen 
at 0.33\,fm/$c$ at full RHIC energy (increasing smoothly to 1\,fm/$c$ at 
$\sqrt{s}$=20\,GeV) and 0.2\,fm/$c$ in the LHC regime. 
This results in initial temperatures of 225\,MeV and 330\,MeV 
in minimum-bias (MB) Au-Au collisions at 20 and 200\,GeV, respectively, increasing to 
$\sim$380\,MeV in central Au-Au(200GeV), and $\sim$560(620)\,MeV in central Pb-Pb 
at 2.76(5.5)\,ATeV. These values differ slightly from previous calculations with 
a quasiparticle EoS; they are also sensitive to the initial spatial profile, 
cf.~left panel of Fig.~\ref{fig_temp}. However, for our main objective of 
calculating low-mass dilepton spectra the initial temperature has little impact.

%%%%%%%%%%%%%%%%%%%%%%%%%%%%%%%%%%%%%%%%%%%%%%%%%%%%%
\subsection{Nonthermal Sources}
\label{ssec_non-thermal}
%%%%%%%%%%%%%%%%%%%%%%%%%%%%%%%%%%%%%%%%%%%%%%%%%%%%%%
In addition to thermal radiation from the locally equilibrated medium,
dilepton emission in URHICs can arise from interactions prior to 
thermalization (e.g., Drell-Yan annihilation) and from EM decays of 
long-lived hadrons after the fireball has decoupled (e.g., 
Dalitz decays $\pi^0,\eta\to \gamma l^+l^-$
or $\omega,\phi\to l^+l^-$). Furthermore, paralleling the structure in
hadronic spectra, a non-thermal component from hard production will
feed into dilepton spectra, e.g., via Bremsstrahlung from hard partons
traversing the medium~\cite{Turbide:2006mc} or decays of both short- and 
long-lived hadrons which 
have not thermalized with the bulk (e.g., ``hard" $\rho$-mesons or 
long-lived EM final-state decays).
Hadronic final-state decays (including the double semileptonic decay of 
two heavy-flavor hadrons originating from a $c\bar c$ or $b\bar b$ pair
produced together in the same hard process) are commonly referred to as 
the ``cocktail", which is routinely evaluated by the experimental
collaborations using the vacuum properties of each hadron with $p_t$
spectra based on measured spectra, or appropriately extrapolated using 
thermal blast-wave models. In URHICs, the
notion of the cocktail becomes problematic for short-lived resonances
whose lifetime is comparable to the duration of the freezeout process
of the fireball (e.g., for $\rho$, $\Delta$, etc.). In their case a better
approximation is presumably to run the fireball an additional $\sim$1~fm/$c$
to treat their final-decay contribution as thermal radiation including medium
effects. However, care has to be taken in evaluating their dilepton 
$p_t$-spectra, as the latter are slightly different for thermal radiation
and final-state decays (cf.~Ref.~\cite{vanHees:2007th} for a discussion
and implementation of this point).   
For light hadrons at low $p_t$, the cocktail scales with the total number 
of charged particles, $N_{\rm ch}$, at given collision energy and 
centrality, while for hard processes a collision-number scaling 
$\propto N_{\rm coll}$ is in order (and compatible with experiment 
where measured, modulo the effects of ``jet quenching").
The notion of ``excess dileptons" is defined as any additional radiation 
observed over the cocktail, sometimes quantified as an ``enhancement
factor" in a certain invariant-mass range. The excess radiation is then
most naturally associated with thermal radiation, given the usual 
limitation where hard processes take over, i.e., $M,q_t\lsim$~2-3\,GeV.

%%%%%%%%%%%%%%%%%%%%%%%%%%%%%%%%%%%%%%%%%%%%%%%%%%%%%%
\subsection{RHIC-200}
\label{ssec_rhic200}
%%%%%%%%%%%%%%%%%%%%%%%%%%%%%%%%%%%%%%%%%%%%%%%%%%%%%%
We start our discussion of low-mass dilepton spectra at full RHIC energy 
where most of the current experimental information at collider energies is 
available, from both PHENIX~\cite{Adare:2009qk} and STAR~\cite{Zhao:2011wa} 
measurements.

%%%%%%%%%%%%%%%%%%%%%%%%%%%%%%%%%%%%%%%%%%%%%%%%%%%%%
\subsubsection{Invariant-Mass Spectra}
\label{sssec_rhic-m}
%%%%%%%%%%%%%%%%%%%%%%%%%%%%%%%%%%%%%%%%%%%%%%%%%%%%%%
\begin{figure}[!t]
\begin{minipage}{0.5\linewidth}
  \includegraphics[width=0.9\textwidth]{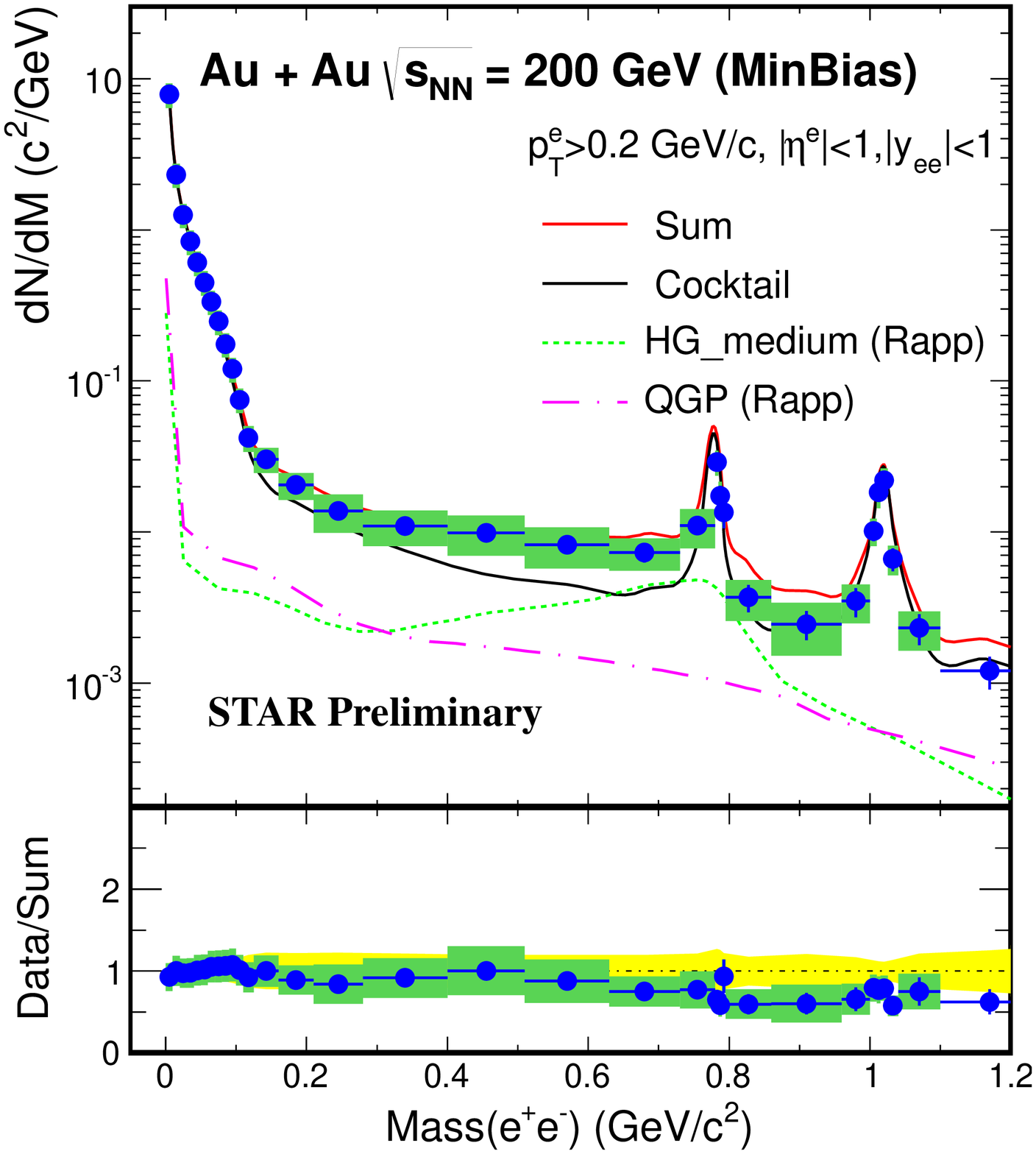}
\end{minipage}
\begin{minipage}{0.5\linewidth}
  \includegraphics[width=0.9\textwidth]{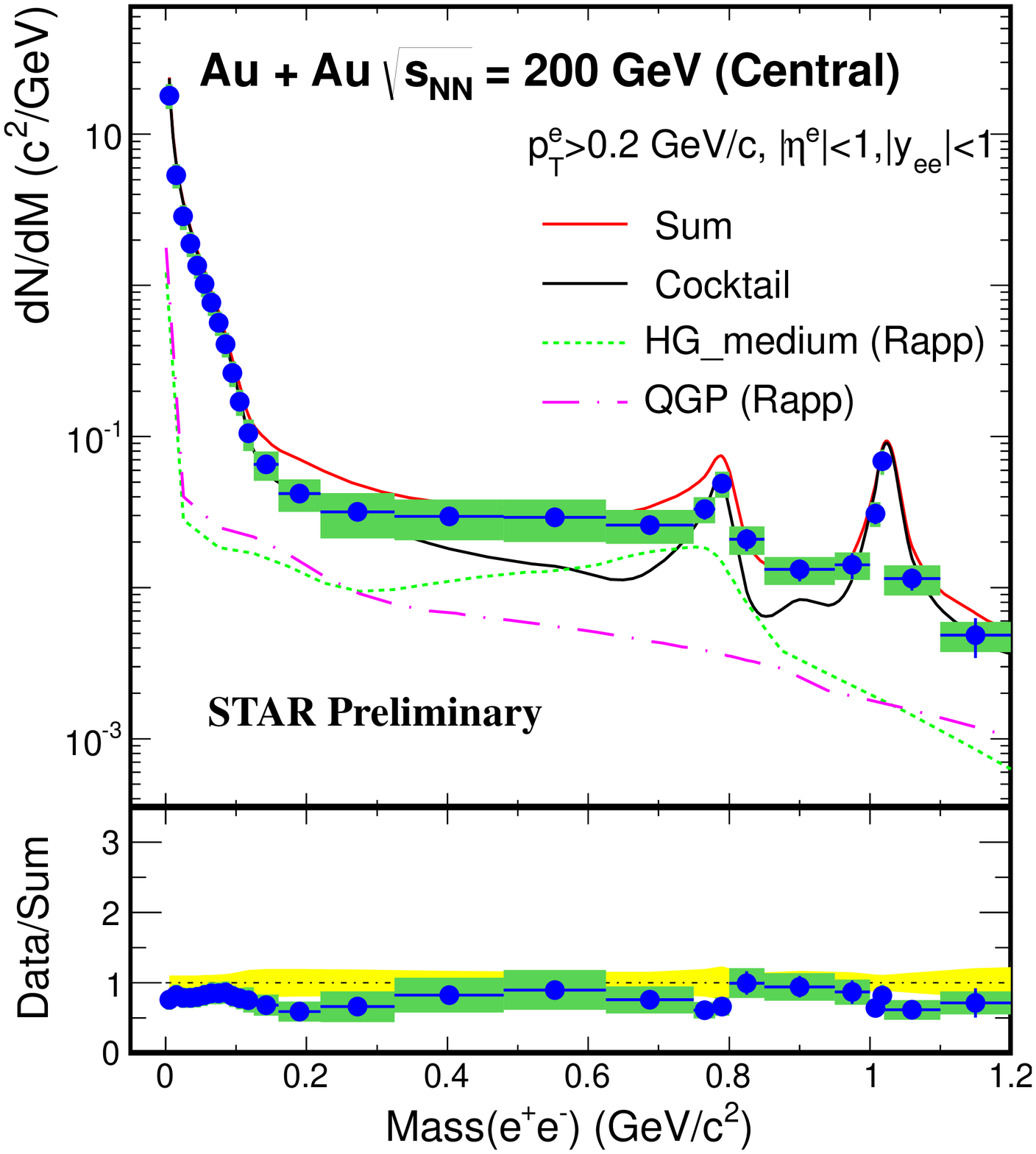}
\end{minipage}
\caption{Dilepton invariant-mass spectra in Au-Au(200\,AGeV) for minimum-bias (left) 
and central (right) collisions.
Theoretical calculations for thermal radiation from a nonperturbative
QGP and in-medium hadronic spectral functions are compared to STAR 
data~\cite{Zhao:2011wa,Zhao:2012}.}
\label{fig_star200}
\end{figure}
Figure~\ref{fig_star200} shows the comparison of thermal fireball 
calculations with low-mass spectra from STAR~\cite{Zhao:2011wa}. As 
compared to earlier calculations with a bag-model EoS~\cite{Rapp:2000pe}, 
the use of lQCD-EoS and emission rates for the QGP enhances the pertinent 
yield significantly. 
It is now comparable to the in-medium hadronic contribution for masses
below $M\simeq0.3$\,GeV, and takes over in the intermediate-mass region 
($M\gsim1.1$\,GeV). The hadronic part of the thermal yield remains  
prevalent in a wide range around the free $\rho$ mass, with a
broad resonance structure and appreciable contributions from 
$4\pi$ annihilation for $M\gsim 0.9$\,GeV. Upon adding the thermal yield
to the final-state decay cocktail by the STAR collaboration (without
$\rho$ decay), the MB data are well described. For the
central data, a slight overestimate around $M\simeq0.2$\,GeV and
around the $\omega$ peak is found. A similar description~\cite{Vujanovic:2012nq} 
of the STAR data arises in a viscous hydrodynamic description of 
the medium using the $\rho$ spectral function from on-shell scattering 
amplitudes~\cite{Eletsky:2001bb} (see also Ref.~\cite{Xu:2011tz}), 
and in the parton-hadron string dynamics transport approach using a schematic 
temperature- and density-dependent broadening in a Breit-Wigner approximation 
of the $\rho$ spectral function~\cite{Linnyk:2011vx}. More studies are
needed to discern the sensitivity of the data to the in-medium spectral
shape, as the latter significantly varies in the different approaches. 
For the PHENIX data (not shown), the enhancement as recently reported in 
Ref.~\cite{Tserruya:2012} for non-central collisions (carried out with the 
hadron-blind detector (HBD) upgrade) agrees 
with earlier measurements~\cite{Adare:2009qk}, 
is consistent with the STAR data, and thus should also agree with theory. 
For the most central Au-Au data, however, a large enhancement was reported 
in Ref.~\cite{Adare:2009qk} which is well above theoretical calculations
with broad spectral 
functions~\cite{Rapp:2000pe,Dusling:2007su,Linnyk:2011vx,Ghosh:2010wt},
even in the MB sample. More ``exotic" explanations of this effect, which did
not figure at the SPS, e.g., a Bose-condensed like glasma in the 
pre-equilibrium stages~\cite{Chiu:2012ij}, have been put forward to explain 
the ``PHENIX puzzle".
However, it is essential to first resolve the discrepancy on the experimental
side, which is anticipated with the HBD
measurement for central collisions.   

To quantify the centrality dependence of the thermal radiation (or excess) yield, 
one commonly introduces an exponent $\alpha_c$ as 
$Y_M(N_{\rm ch})/N_{\rm ch} = C N_{\rm ch}^{\alpha_c}$, which describes how
the excess (or thermal) yield in a given mass range scales
relative to the charged-particle multiplicity. For full RHIC energy, the
theoretical calculation gives $\alpha_c\simeq0.45$ (with a ca.~10\% error),
similar to what had been found for integrated thermal photon 
yields~\cite{Turbide:2003si}.

%%%%%%%%%%%%%%%%%%%%%%%%%%%%%%%%%%%%%%%%%%%%%%%%%%%%%
\subsubsection{Transverse-Momentum Dependencies}
\label{sssec_rhic-qt}
%%%%%%%%%%%%%%%%%%%%%%%%%%%%%%%%%%%%%%%%%%%%%%%%%%%%%%

When corrected for acceptance, invariant-mass spectra are unaffected by any
blue-shift of the expanding medium, which renders them a pristine probe for 
in-medium spectral modifications. However, the different collective flow associated
with different sources may be helpful in discriminating them by investigating 
their $q_t$ spectra, see, e.g., 
Refs.~\cite{Siemens:1985nj,Kajantie:1986dh,Cleymans:1986na,Kampfer:2000nc,Renk:2006qr,Alam:2009da,Deng:2010pq}. 
As is well-known from the observed final-state hadron 
spectra, particles of larger mass experience a larger
blue-shift than lighter particles due to collective motion with the expanding 
medium. Schematically, this can be represented by an effective slope parameter
which for sufficiently large masses takes an approximate form of 
$T_{\rm eff} = T + M \bar{\beta}^2$ where $T$ and $\bar\beta$ denote the local 
temperature and average expansion velocity of the emitting source cell.
Dileptons are well suited to systematically scan the mass dependence of 
$T_{\rm eff}$ by studying $q_t$ 
spectra for different mass bins (provided the data have sufficient statistics). 
At the SPS this has been done by the NA60 collaboration~\cite{Arnaldi:2008fw}, 
who found a gradual increase in the slope from the dimuon threshold to the 
$\rho$ mass characteristic for a source of hadronic origin (aka in-medium 
$\rho$ mesons), a maximum around the $\rho$ mass (late $\rho$ decays), 
followed by a decrease and leveling off in the intermediate-mass region 
(IMR, $M\ge 1$\,GeV) indicative for early emission at temperatures 
$T\simeq$~170-200\,MeV (where at the SPS the collective flow is still small).

\begin{figure}[!t]
\begin{minipage}{0.7\linewidth}
 \includegraphics[width=0.8\textwidth,angle=-90]{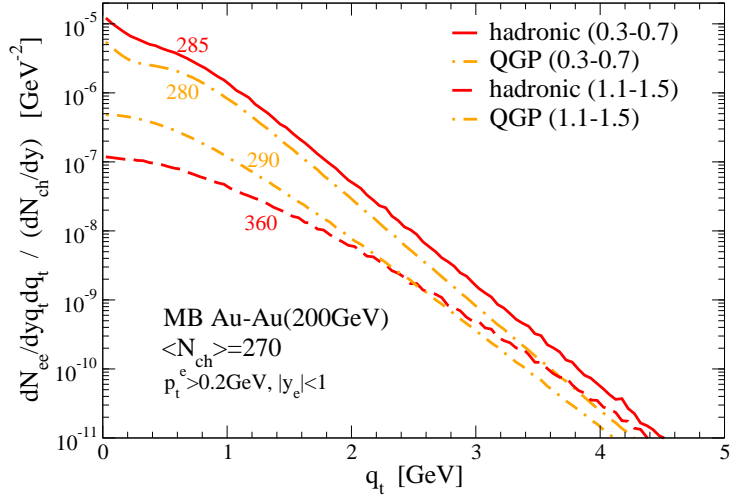}
\end{minipage}
%\begin{minipage}{0.5\linewidth}
%  \includegraphics[width=0.9\textwidth]{dndm-star200cen.eps}
%\end{minipage}
\caption{Dilepton transverse-momentum spectra from
thermal radiation of QGP and hadronic matter in MB Au-Au(200\,AGeV) collisions.
The numbers next to each curve indicate the effective slope parameter, $T_{\rm eff}$
[MeV], as extracted from a two parameter fit using
$dN/(q_tdq_t) = C \exp[-M_t/T_{\rm eff}]$~\cite{vanHees:2007th} with the transverse
mass $M_t=\sqrt{M^2+q_t^2}$ and an average mass of 0.5\,GeV and 1.25\,GeV for
the low- and intermediate-mass window, respectively.}
\label{fig_qt200}
\end{figure}
Figure~\ref{fig_qt200} shows the $q_t$ spectra for thermal radiation from hadronic
matter and QGP in MB Au-Au(200\,AGeV) in two typical mass regions where either of the 
two sources dominates. In the low-mass region (LMR), both sources have a surprisingly
similar slope ($T_{\rm slope}\simeq$~280-285\,MeV), reiterating that the emission 
is from mostly around $T_{\rm pc}$ where the slope of both sources is comparable 
(also recall from Fig.~\ref{fig_star200} that in the mass window $M$=0.3-0.7\,GeV 
the QGP emission is largest at the lower mass end, while the hadronic one is more 
weighted toward the higher end). For definiteness, assuming $T$=170\,MeV and 
$M$=0.5\,GeV, one finds $\bar\beta\simeq$~0.45-0.5, which is right in the expected 
range~\cite{He:2011zx}.
On the other hand, in the IMR, where the QGP dominates, 
the hadronic slope has significantly increased to ca.~360\,MeV due to the larger
mass in the collective-flow term. On the other hand, the slope of the QGP emission 
has only slightly increased over the LMR, indicating that the increase in mass
in the flow-term is essentially offset by an earlier emission temperature, 
as expected for higher mass (for hadronic emission, the temperature
is obviously limited by $T_{\rm pc}$). Consequently, at RHIC the effective 
slope of the total thermal radiation in the IMR exceeds 
the one in the LMR, contrary to what has been observed at SPS. Together with 
blue-shift free temperature measurements from slopes in invariant-mass spectra, this 
provides a powerful tool for disentangling collective and thermal properties
through EM radiation from the medium. 

\begin{figure}[!t]
\begin{minipage}{0.85\linewidth}
\hspace{-0.8cm}
\includegraphics[width=0.9\textwidth,angle=-90]{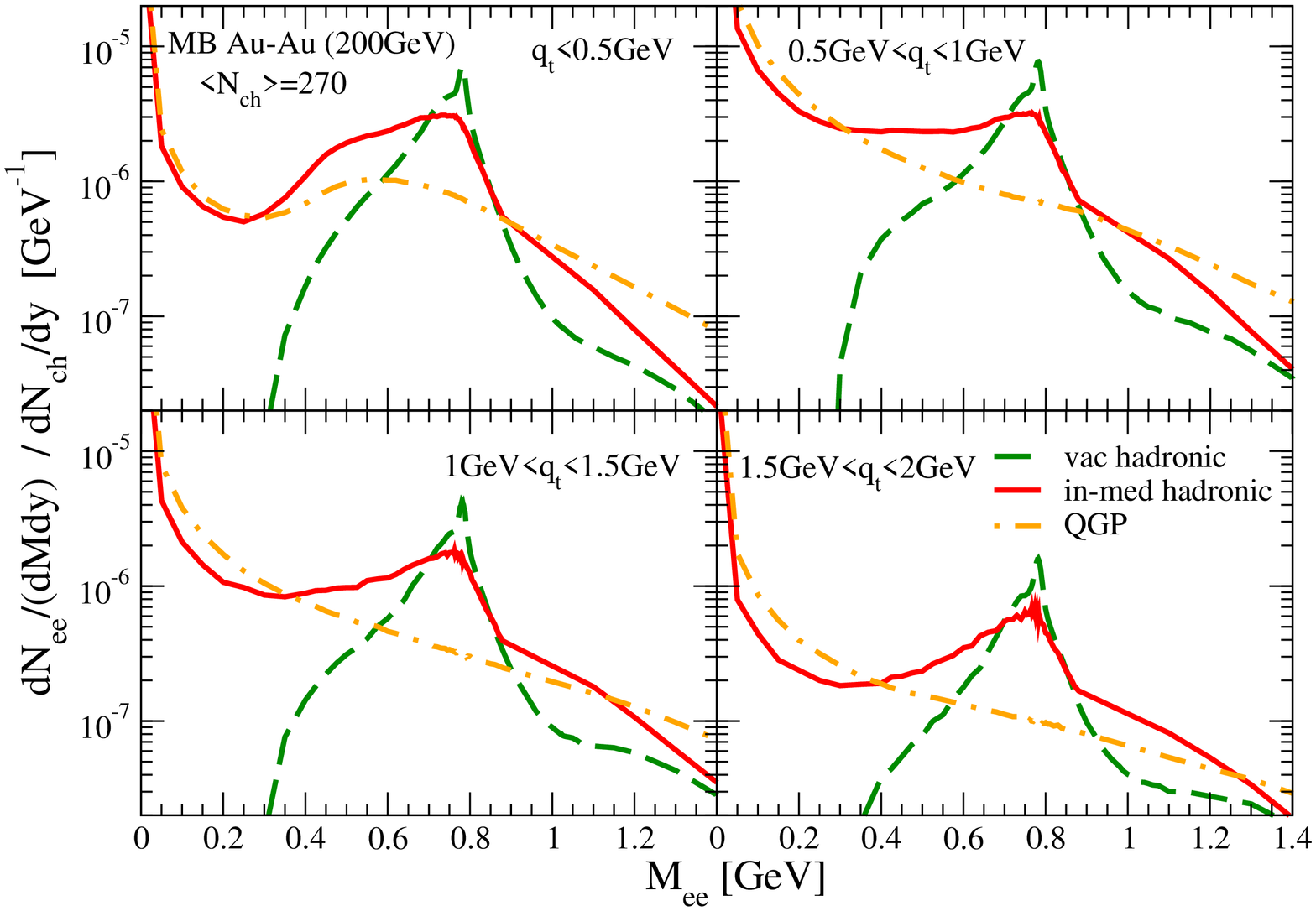}
\end{minipage}
%\begin{minipage}{0.5\linewidth}
%  \includegraphics[width=0.9\textwidth]{dndm-star200cen.eps}
%\end{minipage}
\caption{Dilepton invariant mass-spectra in different bins of transverse momentum
from thermal radiation of QGP (dash-doted line) and hadronic matter (solid line:
in-medium, dashed line: vacuum spectral function) in MB Au-Au(200\,AGeV) collisions;
experimental acceptance as in Figs.~\ref{fig_star200} and \ref{fig_qt200}.}
\label{fig_Mqt200}
\end{figure}
Alternatively, one can investigate the mass spectra in different momentum bins,
possibly revealing a $q_t$-dependence of the spectral shape, as was done for both 
$e^+e^-$ data in Pb-Au~\cite{Agakichiev:2005ai} and $\mu^+\mu^-$ in 
In-In~\cite{Arnaldi:2008fw} at SPS. Calculations for thermal radiation in Au-Au at 
full RHIC energy are shown in Fig.~\ref{fig_Mqt200} for four bins from $q_t$=0-2\,GeV.
One indeed recognizes that the $\rho$ resonance structure becomes more
pronounced as transverse momentum is increased. In the lowest bin the minimum 
structure around $M\simeq0.2$\,GeV is caused by the experimental acceptance, 
specifically the single-electron $p_t^e>$~0.2\,GeV, which for vanishing $q_t$ 
suppresses all dilepton yields below $M\simeq$~$2p_t^{e,min}=$~0.4\,GeV.

%%%%%%%%%%%%%%%%%%%%%%%%%%%%%%%%%%%%%%%%%%%%%%%%%%%%%
\subsubsection{Elliptic Flow}
\label{sssec_rhic-v2}
%%%%%%%%%%%%%%%%%%%%%%%%%%%%%%%%%%%%%%%%%%%%%%%%%%%%%%
\begin{figure}[!t]
\begin{minipage}{0.5\linewidth}
\hspace{-0.6cm}
\vspace{0.2cm}
  \includegraphics[width=1.15\textwidth]{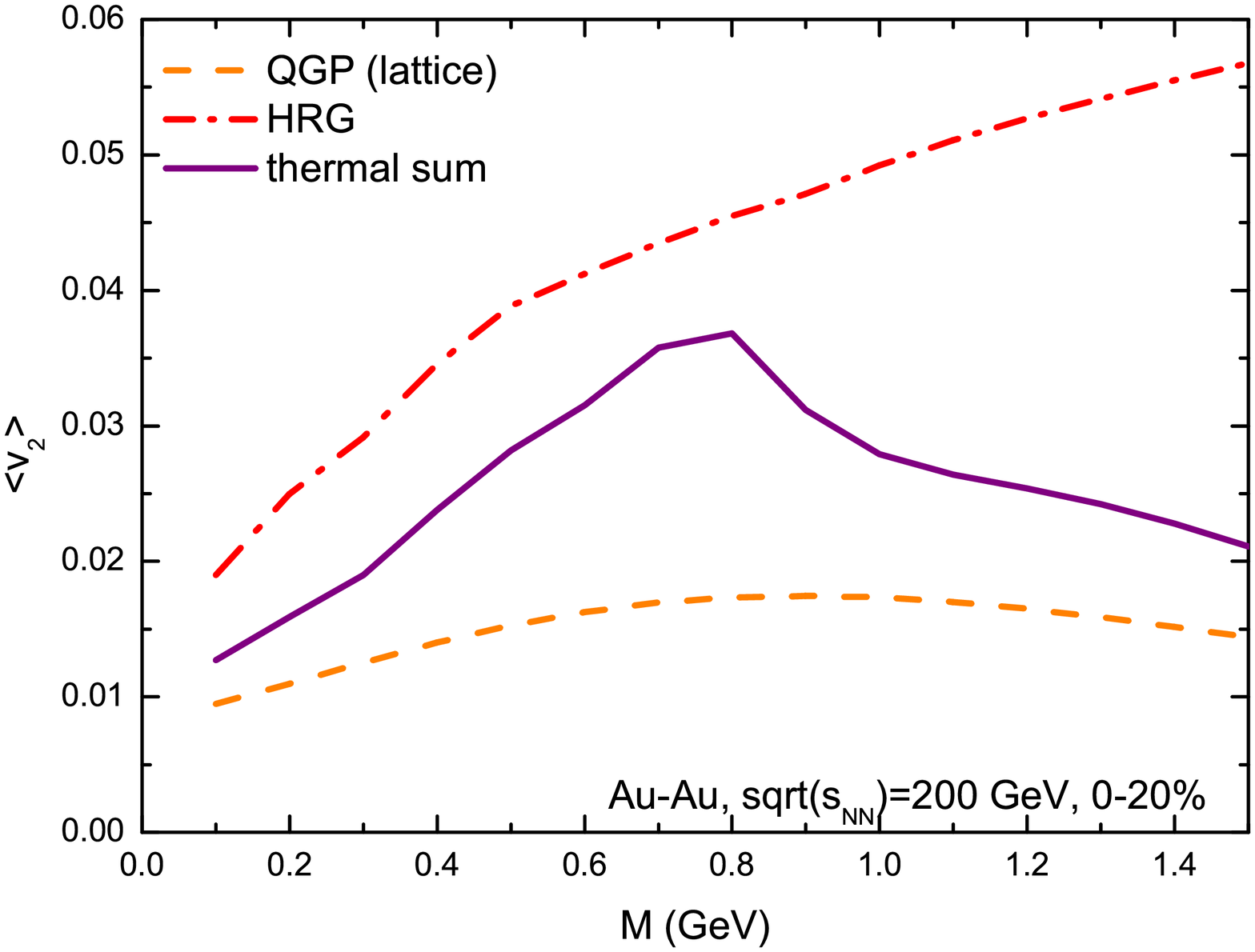}
\end{minipage}
\hspace{-0.3cm}
\begin{minipage}{0.5\linewidth}
  \includegraphics[width=1.0\textwidth]{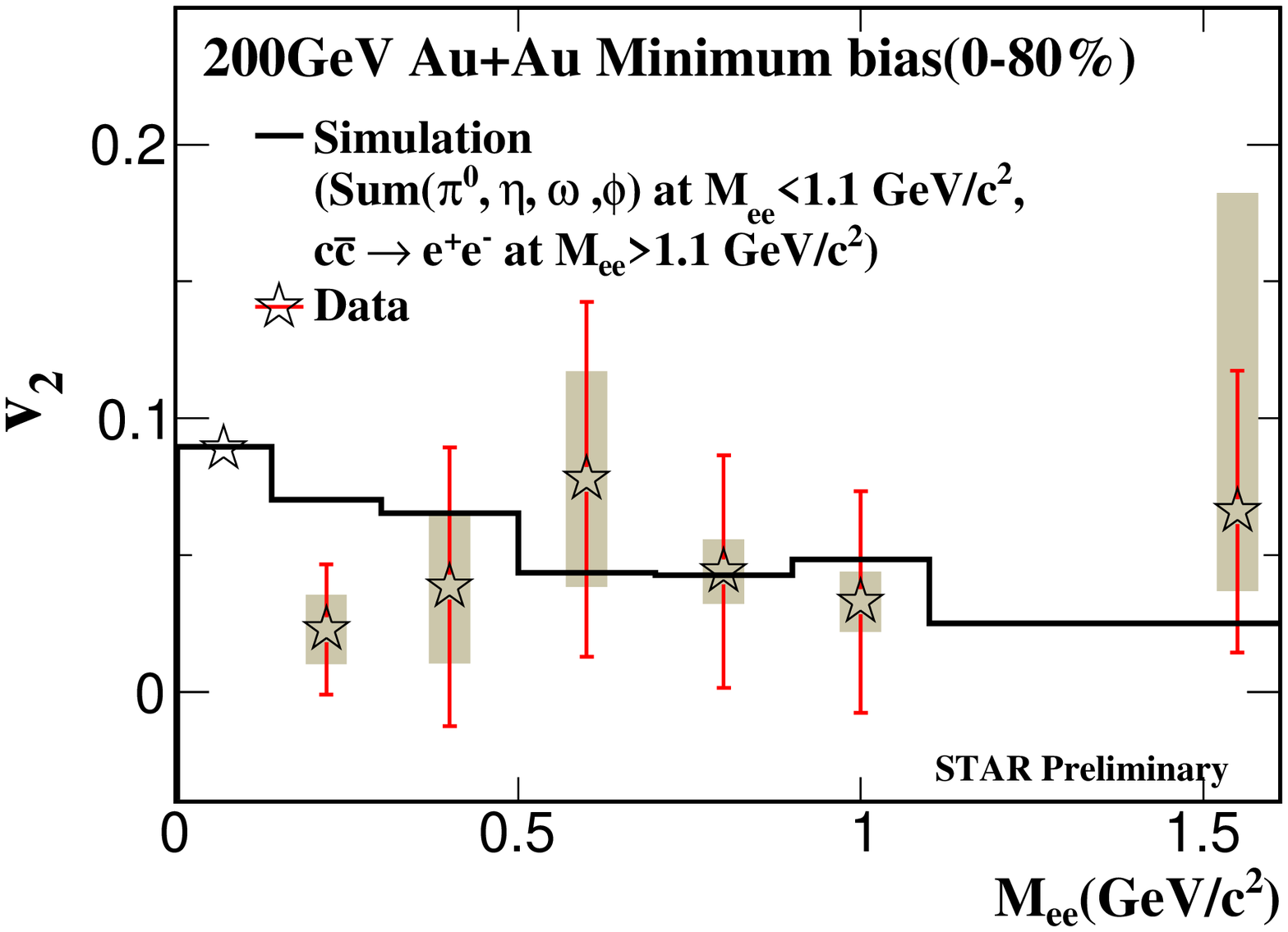}
\end{minipage}
\caption{Left panel: Inclusive elliptic flow of thermal 
dileptons in 0-20\% central Au-Au(200\,AGeV) collisions, calculated within an ideal 
hydrodynamic model with lattice EoS using lQCD-based QGP and medium-modified 
hadronic rates~\cite{He:2013}. Right panel: Dielectron-$v_2$ measured
by STAR in MB Au-Au~\cite{Geurts:2012rv}, including the cocktail contribution; 
the latter has been simulated by STAR and is shown separately by the solid 
histogram.}
\label{fig_v2}
\end{figure}
Another promising observable to diagnose the collectivity, and thus the origin
of the EM emission source, is its elliptic
flow~\cite{Chatterjee:2007xk,Deng:2010pq,Mohanty:2011fp}. The latter is particularly 
useful to discriminate early from late(r) thermal emission sources: contrary to 
the slope parameter, which is subject to an interplay of decreasing temperature and 
increasing flow, the medium's ellipticity is genuinely small (large) in the early 
(later) phases.  The left panel of Fig.~\ref{fig_v2} shows hydrodynamic calculations 
of the inclusive thermal dilepton $v_2$ as a function of invariant mass (using the same
emission rates and EoS as in the previous figures)~\cite{He:2013}. One nicely 
recognizes
a broad maximum structure around the $\rho$ mass, indicative for predominantly
later emission source in the vicinity of its vacuum mass, a characteristic mass
dependence (together with an increasing QGP fraction) below, and a transition
to a dominant QGP fraction with reduced $v_2$ above. All these features are
essentially paralleling the mass dependence of the {\em slope} parameter
at SPS, while the latter exhibits a marked increase at RHIC in the IMR due to the 
increased radial flow in the QGP and early hadronic phase. Rather similar results
are obtained in  hydrodynamic calculations with in-medium spectral
functions from on-shell scattering amplitudes~\cite{Vujanovic:2012nq}.  When
using a less pronounced in-medium broadening, the peak structure in $v_2(M)$
tends to become narrower~\cite{Chatterjee:2007xk,Deng:2010pq,Mohanty:2011fp}.
First measurements of the dilepton-$v_2$ have been presented by 
STAR~\cite{Geurts:2012rv}, see right panel of Fig.~\ref{fig_v2}. The shape
of the data is not unlike the theoretical calculations, while it is also
consistent with the simulated cocktail contribution. Note that the total 
$v_2$ is essentially a weighted sum of cocktail and excess radiation. Thus, 
if the total $v_2$ were to agree with the cocktail, it implies that the $v_2$ of 
the excess
radiation is as large as that of the cocktail. Clearly, future $v_2$ 
measurements with improved accuracy will be a rich source of information.   

Significant $v_2$ measurements of EM excess radiation have recently been reported 
in the $M$=0 limit, i.e., for direct photons, by both 
PHENIX~\cite{Adare:2008fq,Adare:2011zr} and ALICE~\cite{Wilde:2012wc,Lohner:2012ct}.  
A rather large $v_2$ signal has been observed in both 
experiments~\cite{Adare:2011zr,Lohner:2012ct}, suggestive for rather late 
emission~\cite{vanHees:2011vb} (see also 
Refs.~\cite{Liu:2009kta,Holopainen:2011pd,Dion:2011pp,Basar:2012bp}).
In addition, the effective slope parameters of the excess radiation have
been extracted, $T_{\rm eff}= 219\pm27$\,MeV~\cite{Adare:2008fq} at RHIC-200 and
$304\pm51$\,MeV at LHC-2760~\cite{Wilde:2012wc}, which are rather soft 
once blue-shift effects are accounted for. In fact, these slopes are not 
inconsistent with the trends in the LMR dileptons when going from RHIC 
(Fig.~\ref{fig_qt200} above) to LHC (Fig.~\ref{fig_qt5500} below). This would 
corroborate their main origin from around $T_{\rm pc}$.

%%%%%%%%%%%%%%%%%%%%%%%%%%%%%%%%%%%%%%%%%%%%%%%%%%%%%%
\subsection{RHIC Beam Energy Scan}
\label{ssec_rhic-bes}
%%%%%%%%%%%%%%%%%%%%%%%%%%%%%%%%%%%%%%%%%%%%%%%%%%%%%%
A central question for studying QCD phase structure is how the spectral properties
of excitations behave as a function of chemical potential and temperature. With
the EM (or vector) spectral function being the only one directly accessible via
dileptons, systematic measurements as a function of beam energy are mandatory. 
At fixed target energies, this is being addressed by the current and future HADES 
efforts ($E_{\rm beam}$=1-10\,AGeV)~\cite{Agakishiev:2009yf,Holzmann:2012pa}, 
by CBM for $E_{\rm beam}$(Au) up to $\sim$35\,AGeV~\cite{Friman:2011zz}, and has 
been measured at SPS energies at $E_{\rm beam}$=158\,AGeV, as well as in a CERES 
run at 40\,AGeV~\cite{Adamova:2002kf}.    

\begin{figure}[!t]
\begin{minipage}{0.9\linewidth}
  \includegraphics[width=1.0\textwidth]{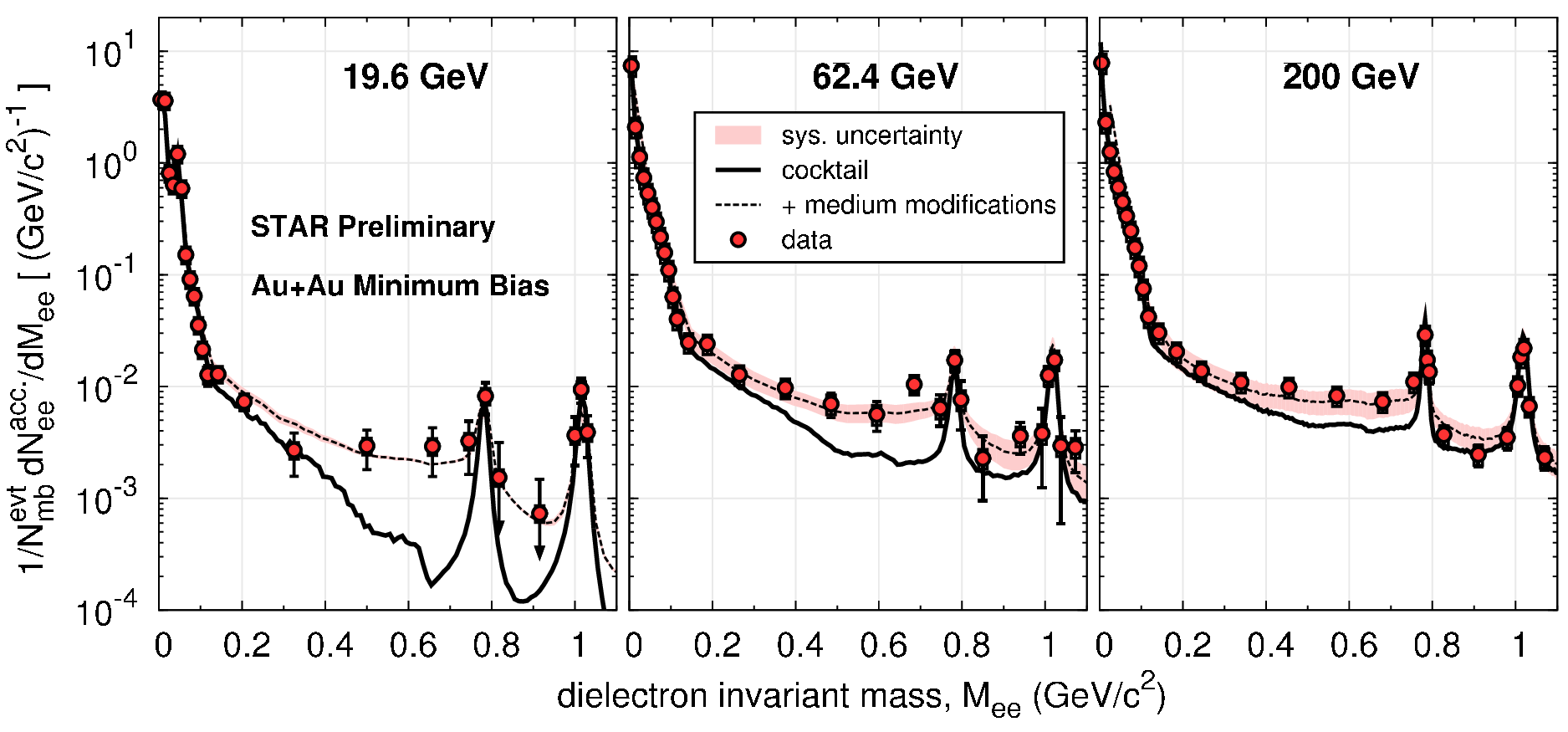}
\end{minipage}
\caption{Low-mass dilepton spectra as measured by STAR in the RHIC beam-energy
scan~\cite{Geurts:2012rv}; MB spectra are compared to theoretical predictions
for the in-medium hadronic + QGP radiation, added to the cocktail contribution.}
\label{fig_bes}
\end{figure}
At collider energies, a first systematic study of the excitation function of
dilepton spectra has been conducted by STAR~\cite{Geurts:2012rv} as part of the 
beam-energy scan program at RHIC. The low-mass excess radiation develops 
smoothly when going down form $\sqrt{s_{NN}}$=200\,GeV via 62\,GeV to 20\,GeV, 
cf.~Fig.~\ref{fig_bes}. Closer inspection reveals that the enhancement factor
of excess radiation over cocktail in the region below the $\rho$ mass
increases as the energy is reduced~\cite{Geurts:2012rv}. An indication of a similar
trend was observed when comparing the CERES measurements in Pb-Au at 
$\sqrt{s_{NN}}$=17.3\,GeV and 8.8\,GeV. Theoretically, this can be understood
by the importance of baryons in the generation of medium effects~\cite{Rapp:1997fs}, 
specifically the low-mass enhancement in the $\rho$ spectral function. These 
medium effects become stronger as the beam energy is reduced since the hadronic 
medium close to $T_{\rm pc}$ becomes increasingly baryon rich. At the same time, 
the cocktail contributions, which are mostly made up by meson decays, decrease. 
The hadronic in-medium effects are expected to play a key role in the 
dilepton excess even at collider energies. The comparison with the STAR 
excitation function supports the interpretation of the excess radiation as 
originating from a melting $\rho$ resonance in the vicinity of $T_{\rm pc}$.       

A major objective of the beam-energy program is the search for a critical point.
One of the main effects associated with a second-order endpoint is the critical
slowing down of relaxation rates due to the increase in the correlation length
in the system. For the medium expansion in URHICs, this may imply an ``anomalous" 
increase in the lifetime of the interacting fireball. If this is so, dileptons 
may be an ideal tool to detect this phenomenon, since their total yield (as 
quantified by their enhancement factor) is directly proportional to the duration 
of emission. The NA60 data have shown that such a lifetime measurement 
can be carried out with an uncertainty of about $\pm$1\,fm/$c$~\cite{vanHees:2007th}. 
In the calculations shown in Fig.~\ref{fig_bes} no critical slowing down has been
assumed; as a result, the average lifetime in MB Au-Au collisions increases
smoothly from ca.~8 to 10\,fm/$c$. Thus, if a critical point were to exist and
lead to a, say, 30\% increase in the lifetime in a reasonably localized range
of beam energies, dilepton yields ought to be able to detect this signature.
This signal would further benefit from the fact that the prevalent radiation
arises from around $T_{\rm pc}$ where the largest effect from the slowing down 
is expected.  

%%%%%%%%%%%%%%%%%%%%%%%%%%%%%%%%%%%%%%%%%%%%%%%%%%%%%%
\subsection{LHC}
\label{ssec_lhc}
%%%%%%%%%%%%%%%%%%%%%%%%%%%%%%%%%%%%%%%%%%%%%%%%%%%%%%
The previous section raises the question whether the smooth excitation 
function of dilepton invariant-mass spectra in the RHIC regime will continue 
to LHC energies, which increase by another factor of $\sim$20. On the 
other hand, the dilepton $q_t$ spectra, especially their inverse-slope 
parameters, indicate an appreciable variation from SPS to RHIC, increasing 
from ca.~220 to 280\,MeV in the LMR, and, more pronounced, from ca.~210 to 
about 320\,MeV in the IMR. This is a direct consequence of the stronger 
(longer) development of collective flow in the QGP phase of the fireball 
evolution. This trend will continue at the LHC, as we will see below. In the 
following section (\ref{sssec_lhc-ee}), we will first discuss the dielectron 
channel at LHC, and highlight the excellent experimental 
capabilities that are anticipated with a planned major upgrade 
program of the ALICE detector~\cite{ALICE-up}. In addition, ALICE can measure 
in the dimuon channel, albeit with somewhat more restrictive cuts whose 
impact will be illustrated in Sec.~\ref{sssec_lhc-mu}.  

%%%%%%%%%%%%%%%%%%%%%%%%%%%%%%%%%%%%%%%%%%%%%%%%%%%%%%
\subsubsection{Dielectrons}
\label{sssec_lhc-ee}
%%%%%%%%%%%%%%%%%%%%%%%%%%%%%%%%%%%%%%%%%%%%%%%%%%%%%%
\begin{figure}[!t]
\begin{minipage}{0.5\linewidth}
\includegraphics[width=0.9\textwidth,angle=-90]{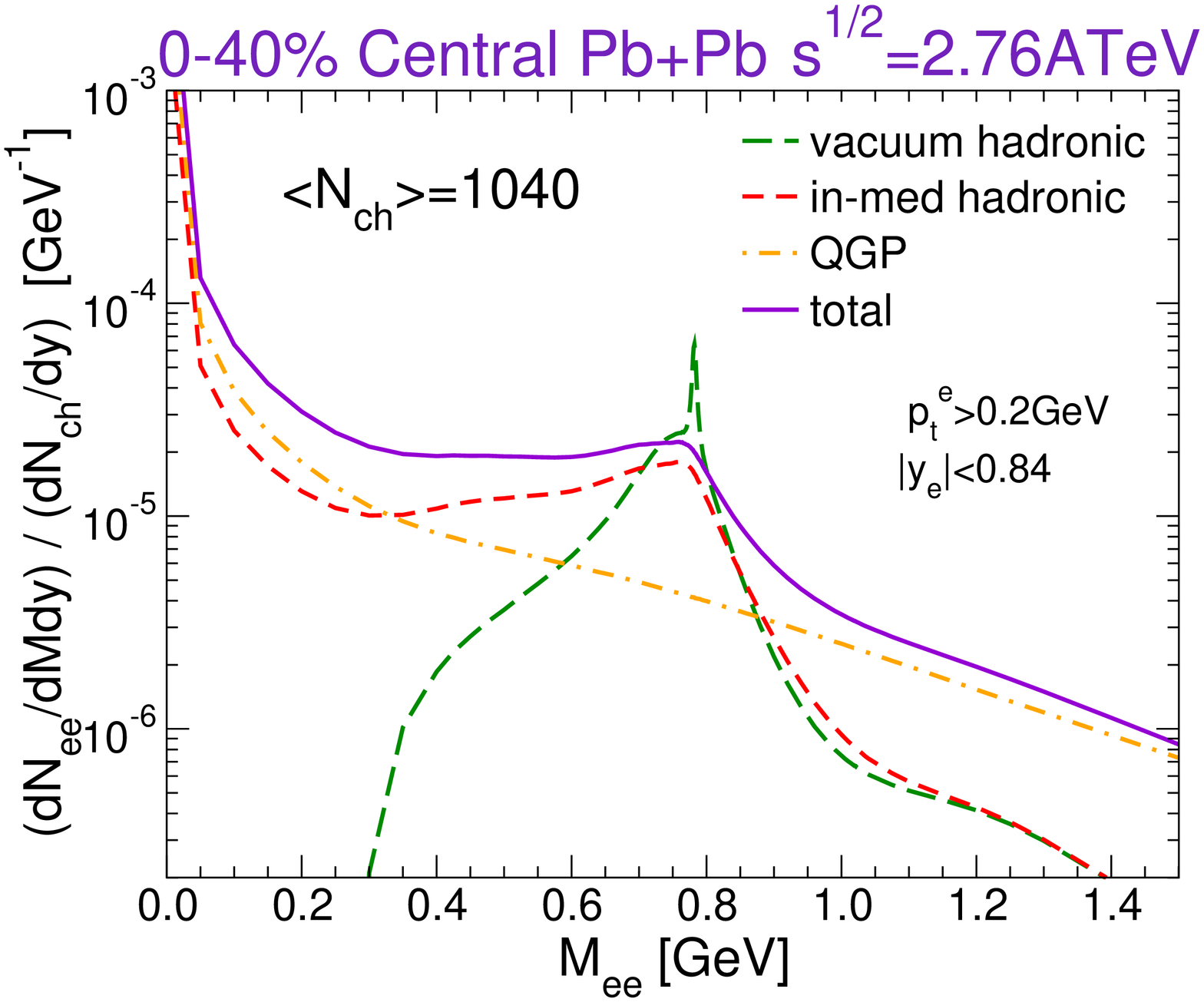}
\end{minipage}
\begin{minipage}{0.5\linewidth}
  \includegraphics[width=0.9\textwidth,angle=-90]{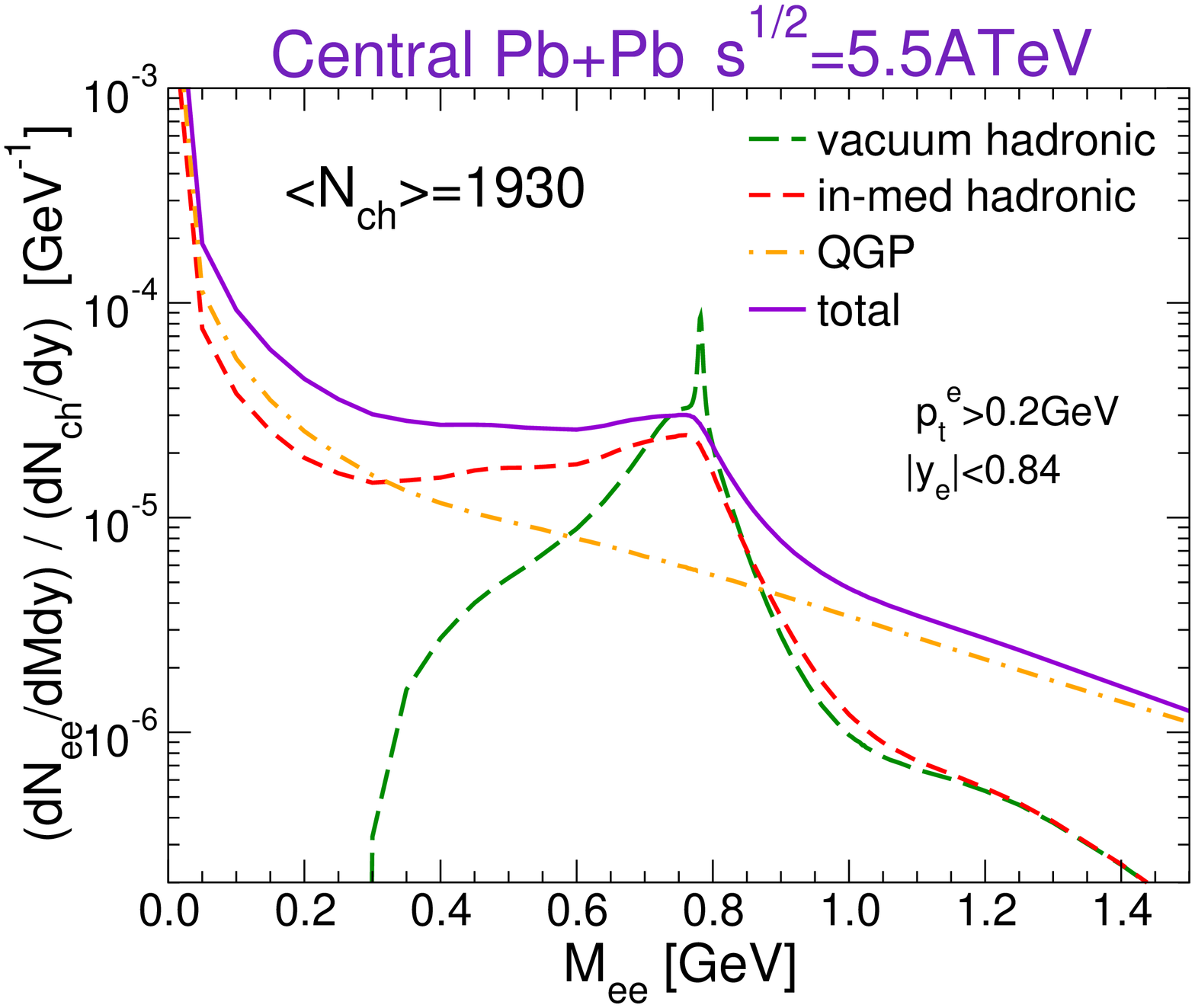}
\end{minipage}
\caption{Dielectron invariant-mass spectra from thermal radiation in 0-40\% 
central Pb-Pb(2.76\,ATeV) (left panel) and 0-10\% central Pb-Pb(5.5\,ATeV) 
(right panel), including single-electron cuts to simulate the ALICE acceptance. 
Hadronic (with in-medium or vacuum EM spectral function) and QGP contributions 
are shown separately along with the sum of in-medium hadronic plus QGP. 
Here and in the following LHC plots, both vacuum and in-medium hadronic 
emission rates in the LMR have been supplemented with the vacuum spectral 
function in the LMR, i.e., no in-medium effects due to chiral mixing have been 
included (for all RHIC calculations shown in the previous sections
full chiral mixing was included).}
\label{fig_M-lhc}
\end{figure}
The invariant-mass spectra of thermal radiation at LHC energies show a 
very similar shape and hadronic/QGP composition as at RHIC energy, see 
Fig.~\ref{fig_M-lhc}. This is not surprising given the virtually 
identical in-medium hadronic and QGP rates 
along the thermodynamic trajectories at RHIC and LHC (where $\mu_B\ll T$ 
at chemical freezeout). It implies that the thermal radiation into the 
LMR is still dominated by temperatures around $T_{\rm pc}$, with 
little (if any) sensitivity to the earliest phases. The total yield, on the 
other hand, increases substantially due to the
larger fireball volumes created by the larger multiplicities. More 
quantitatively, the ($N_{\rm ch}$-normalized) enhancement around, e.g., 
$M$=0.4\,GeV, approximately scales as $N_{\rm ch}^{\alpha_{E}}$ with 
$\alpha_{E}\simeq0.8$ relative to central Au-Au at full RHIC energy. This 
is a significantly stronger increase than the centrality dependent 
enhancement at fixed collision energy,  $\alpha_c\simeq0.45$ as quoted in 
Sec.~\ref{sssec_rhic-m}.   

\begin{figure}[!t]
\begin{minipage}{0.5\linewidth}
  \includegraphics[width=1.0\textwidth]{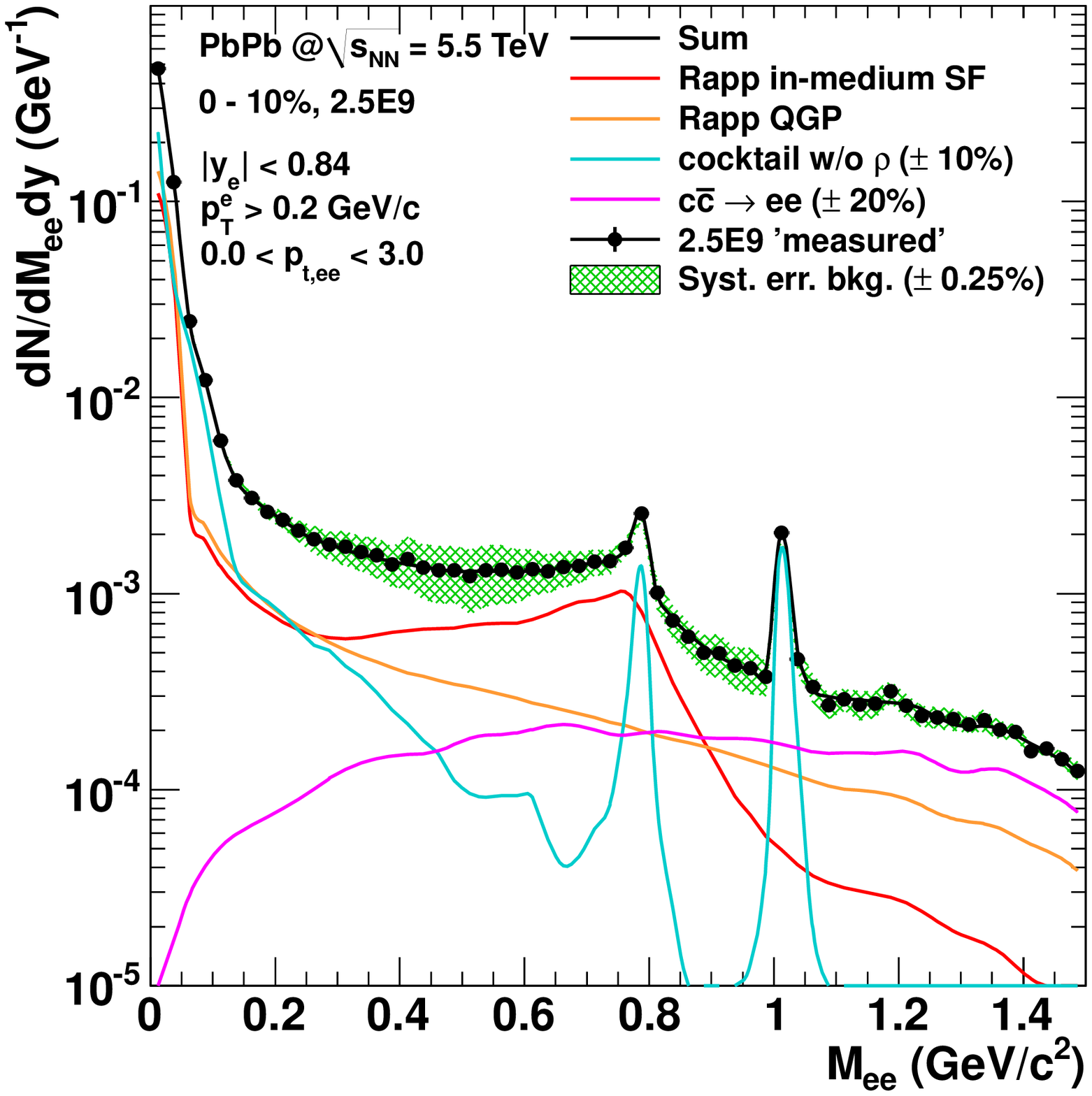}
\end{minipage}
\begin{minipage}{0.5\linewidth}
  \includegraphics[width=1.0\textwidth]{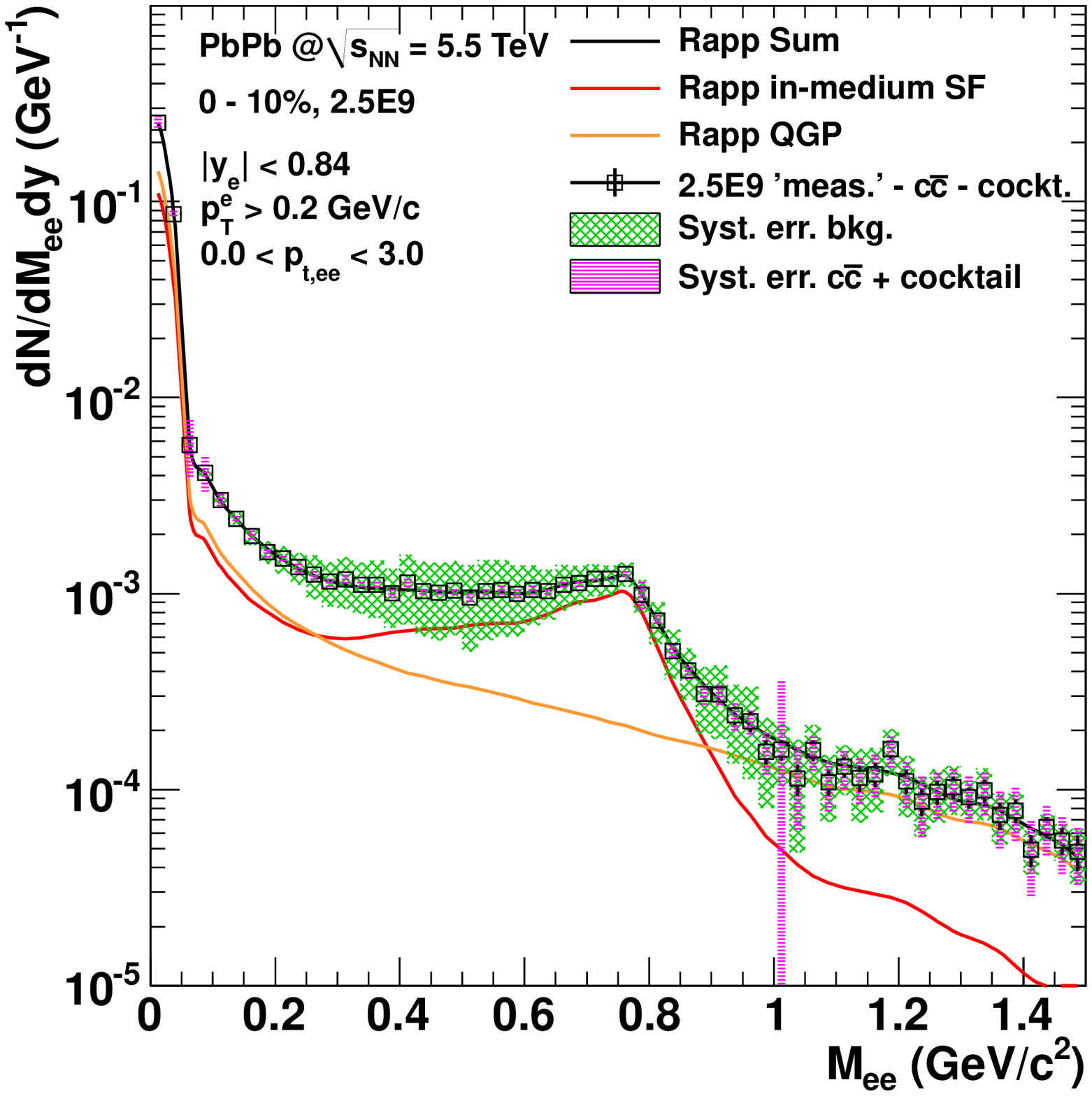}
\end{minipage}
\caption{Simulations of dielectron invariant-mass spectra in Pb-Pb(5.5~ATeV)
collisions assuming the thermal spectra shown in the left panel of
Fig.~\ref{fig_M-lhc} as the excess signal~\cite{ALICE-up,Reichelt:2013}. 
In addition to the acceptance cuts on single-electron rapidity and momentum, 
pair efficiency and displaced vertex cuts are included here. 
Left panel: Invariant-mass spectra
after subtraction of combinatorial background; in addition to the thermal
signal the simulated data contain the hadronic cocktail and correlated
open-charm decays. Right panel: Simulated excess spectra after subtraction
of the cocktail and the open-charm contribution using displaced vertex cuts.
}
\label{fig_sim-lhc}
\end{figure}
Detailed simulation studies of a proposed major upgrade of the ALICE detector 
have been conducted in the context of a pertinent letter of 
intent~\cite{ALICE-up}. The final results after subtraction of uncorrelated
(combinatorial) background are summarized in Fig.~\ref{fig_sim-lhc}, based
on an excess signal given by the thermal contributions in 
Fig.~\ref{fig_M-lhc}.\footnote{The thermal yields provided 
for the simulations were later found to contain a coding error in the author's 
implementation of the experimental acceptance; the error turns out to be rather 
inconsequential for the shape and relative composition of the signal, as a close 
comparison of the right panels in Fig.~\ref{fig_M-lhc} and 
\ref{fig_sim-lhc} reveals; the absolute differential yields differ by up to 
20-30\%.} 
The left panel shows that the thermal signal is dominant for the most
part of the LMR (from ca.~0.2-1.\,GeV), while in the IMR it is outshined 
by correlated heavy-flavor decays. However, the latter can be effectively 
dealt with using displaced vertex cuts, while the excellent mass resolution,
combined with measured and/or inferred knowledge of the Dalitz spectra of 
$\pi^0$ (from charged pions), $\eta$ (from charged kaons) and $\omega$ (from 
direct dilepton decays), facilitate a reliable subtraction of the cocktail. 
The resulting excess spectra shown in the right panel are of a quality 
comparable to the NA60 data. This will allow for quantitative studies of the
in-medium EM spectral function in the LMR which are critical for being able 
to evaluate signatures of chiral restoration (as discussed elsewhere, see, 
e.g., Refs.~\cite{Hohler:2012fj,Rapp:2012zq}). In addition, the yield and
spectral slope of the dominantly QGP emission in the IMR will open
a pristine window on QGP lifetime and temperature (recall that the $M$
spectra, which are little affected by the acceptance cuts in the IMR,
are unaffected by blue shifts). 

\begin{figure}[!t]
\begin{minipage}{0.7\linewidth}
 \includegraphics[width=0.8\textwidth,angle=-90]{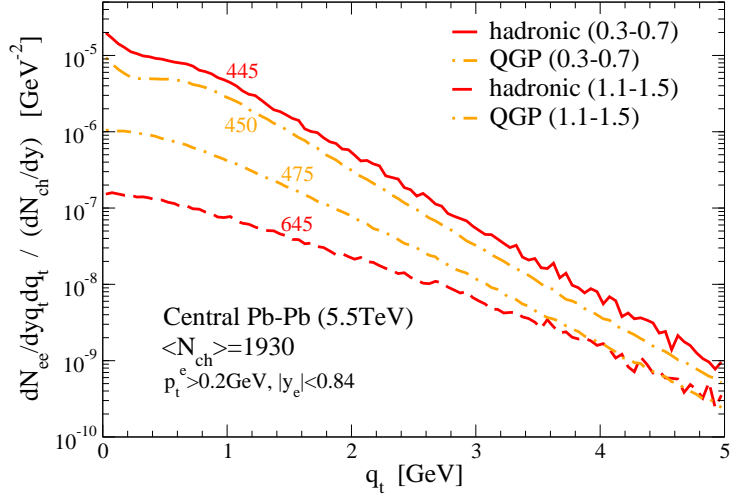}
\end{minipage}
\caption{Same as Fig.~\ref{fig_qt200}, but for
central Pb-Pb(5.5~ATeV).}
\label{fig_qt5500}
\end{figure}
Let us turn to the dilepton $q_t$ spectra at full LHC energy, displayed again 
for two mass bins representing the LMR and IMR in Fig.~\ref{fig_qt5500}. 
Compared to RHIC, 
the LHC fireball is characterized by a marked increase in QGP lifetime 
and associated build-up of transverse flow by the time the system has 
cooled down to $T_{\rm pc}$. Consequently, the $q_t$ spectra exhibit an
appreciable increase in their inverse-slope parameters, by about
60\% in the LMR (for both hadronic and QGP parts) and for the QGP part in the 
IMR, and up to 80\% for the hadronic IMR radiation (recall that in a scenario
with chiral mixing, the hadronic radiation for $M$=1.1-1.5\, GeV is expected to 
increase by about a factor of 2, so that its larger slope compared to the QGP
will become more significant for the total).

%%%%%%%%%%%%%%%%%%%%%%%%%%%%%%%%%%%%%%%%%%%%%%%%%%%%%%
\subsubsection{Dimuons}
\label{sssec_lhc-mu}
%%%%%%%%%%%%%%%%%%%%%%%%%%%%%%%%%%%%%%%%%%%%%%%%%%%%%%
Low-mass dilepton measurements are also possible with ALICE in the 
dimuon channel at forward rapidities, $2.5<y_\mu<4$, albeit with
somewhat more restrictive momentum cuts~\cite{Tieulent:2012}.  
The charged-particle multiplicity in this rapidity range is reduced
by about 30\% compared to midrapidity~\cite{Gulbrandsen:2013iu}, but, 
at 2.76~ATeV, is still ca.~30\% above central rapidities in central Au-Au 
at RHIC. 

\begin{figure}[!t]
\begin{minipage}{0.5\linewidth}
  \includegraphics[width=0.88\textwidth,angle=-90]{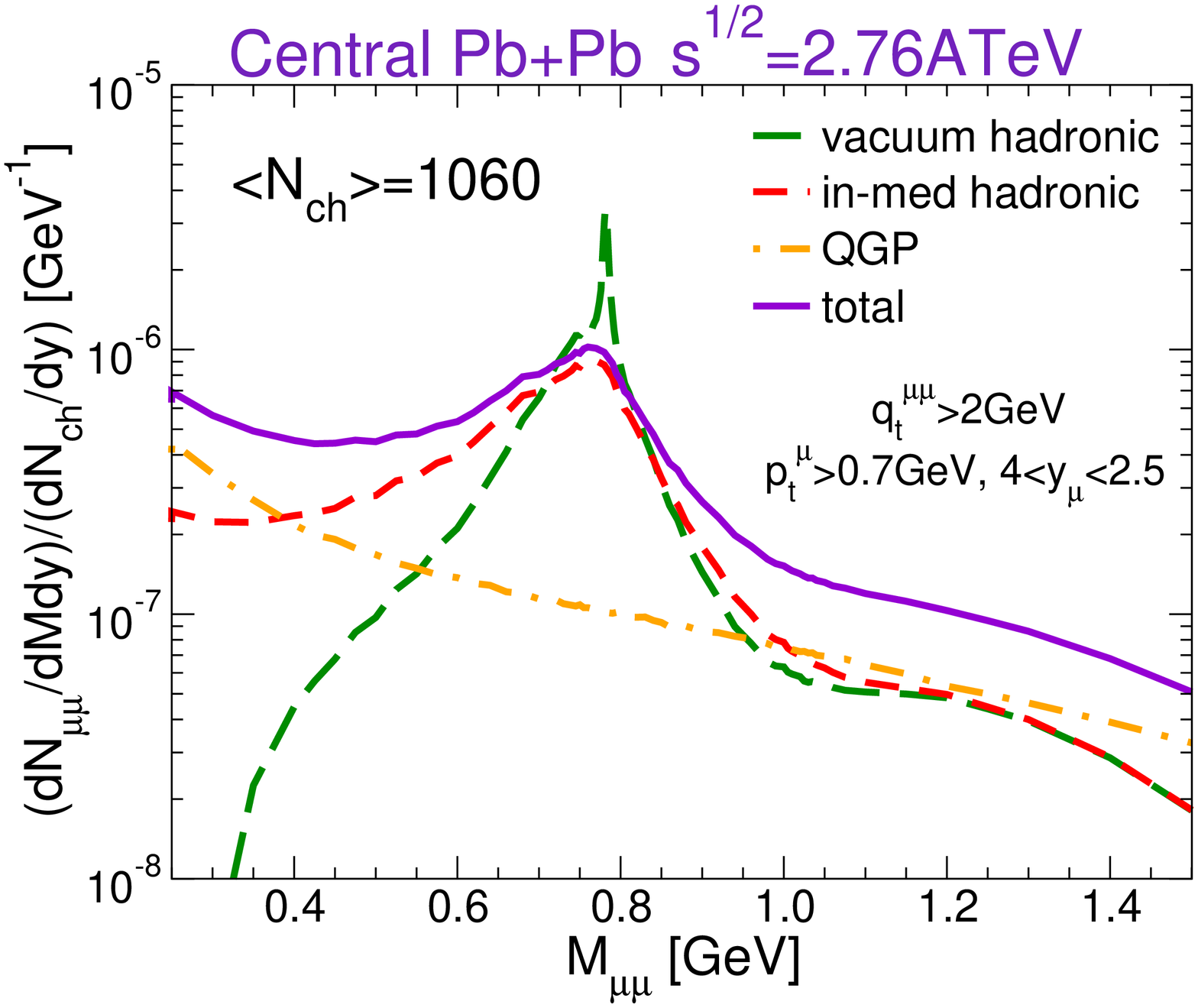}
\end{minipage}
\begin{minipage}{0.5\linewidth}
  \includegraphics[width=0.88\textwidth,angle=-90]{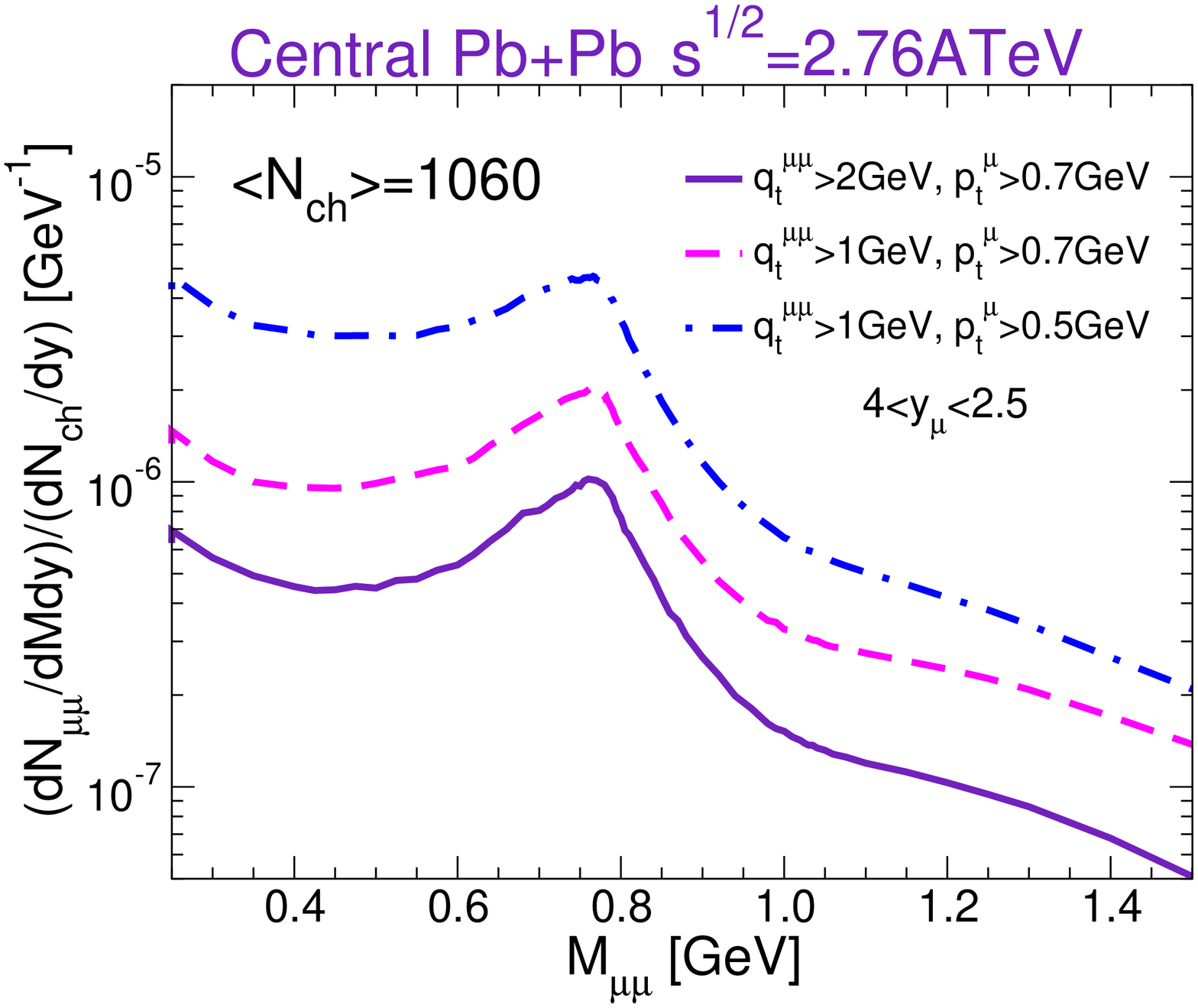}
\end{minipage}
\caption{Calculations of thermal dimuon invariant-mass spectra in central 
Pb-Pb(2.76\,ATeV) collisions at forward rapidity, $y$=2.5-4.
In the left panel, in-medium hadronic, vacuum hadronic, QGP and the sum of
in-medium hadronic plus QGP are shown with ``strong" cuts on single- and dimuon
transverse momenta. The right panel illustrates how the total yield
increases when the two cuts are relaxed.}
\label{fig_lhc-mu}
\end{figure}
Figure~\ref{fig_lhc-mu} illustrates the expected thermal mass spectra in
central Pb-Pb(2.76\,ATeV). For ``conservative" cuts on the di-/muons 
($q_t^{\mu\mu}>2$\,GeV, $p_t^\mu>0.7$\,GeV), their yield is substantially 
suppressed (see left panel), by about one order of magnitude, compared to 
a typical single-$e$ cut of $p_t^e>0.2$\,GeV. In addition, the spectral
broadening of the in-medium $\rho$ meson is less pronounced, a trend that
was also observed in the $q_t$-sliced NA60 dimuon spectra.  Here it
is mostly due to the suppression of medium effects at larger $\rho$-meson
momentum relative to the heat bath, caused by hadronic formfactors (analogous
to RHIC, recall Fig.~\ref{fig_Mqt200}).
It is, in fact, mostly the pair cut which is responsible for the 
suppression, since $q_t^{\mu\mu,\rm cut}$ is significantly larger than 
2$p_t^{\mu,\rm cut}$. If the former can be lowered to, say, 1\,GeV, the 
thermal yield of accepted pairs increases by about a factor 3 in the IMR
and 2 in the LMR (see dashed line in the right panel of Fig.~\ref{fig_lhc-mu}).
The LMR acceptance is now mainly limited by the single-$\mu$ cut, as the 
latter suppresses low-mass pairs whose pair momentum is not at least 
2$p_t^{\mu,\rm cut}$ (the same effect is responsible for the rather sharp
decrease in acceptance for low-momentum electron pairs below $M\simeq0.4$\,GeV 
in the upper left panel of Fig.\ref{fig_Mqt200}, leading to a dip toward
lower mass in the thermal spectra, even though the thermal rate increases 
approximately exponentially). This could be much improved by lowering the
single-$\mu$ cut to, e.g., 0.5\,GeV, which would increase the low-mass yield
by about a factor of 3. At the same time, the spectral broadening of the 
$\rho$ becomes more pronounced in the accepted yields, i.e., the data would 
be more sensitive to medium effects.

%%%%%%%%%%%%%%%%%%%%%%%%%%%%%%%%%%%%%%%%%%%%
\section{Summary and Outlook}
\label{sec_sum}
%%%%%%%%%%%%%%%%%%%%%%%%%%%%%%%%%%%%%%%%%%%%%%%%%%%%%%
In this article we have given an overview of medium modifications of the 
electromagnetic spectral function under conditions expected at collider 
energies (high temperature and small baryon chemical potential), and how 
these medium effects manifest themselves in experimental dilepton spectra at 
RHIC and LHC. 
For the emission rates from the hadronic phase we have focused on 
predictions from effective hadronic Lagrangians evaluated with standard 
many-body (or thermal field-theory) techniques; no in-medium changes of the 
parameters in the Lagrangian (masses and couplings) have been assumed. Since
this approach turned out to describe dilepton data at the SPS well, providing
testable predictions for upcoming measurements at RHIC and LHC is in order. 
As collision energy increases the QGP occupies an increasing 
space-time volume of the fireball evolution. To improve the description of 
the pertinent dilepton radiation, information from lattice-QCD has been 
implemented on (i) the equation of state around and above $T_{\rm pc}$, and 
(ii) nonperturbative dilepton emission rates in the QGP. The latter have 
been ``minimally" extended to finite 3-momentum to facilitate their 
use in calculations of observables. Since these rates are rather 
similar to previously employed perturbative (HTL) rates, an approximate 
degeneracy of the in-medium hadronic and the lQCD rates close to 
$T_{\rm pc}$ still holds. This is welcome in view of the smooth crossover 
transition as deduced from bulk properties and order parameters at $\mu_q$=0. 

The main features of the calculated thermal spectra at RHIC and LHC are as 
follows. The crossover in the lQCD EoS produces a noticeable increase of 
the QGP fraction of the yields (compared to a bag-model EoS), while the 
hadronic portion decreases (its former mixed-phase contribution has been 
swallowed by the QGP). However, due to the near-degeneracy of the QGP and 
hadronic emission rates near $T_{\rm pc}$, both the total yield and its 
spectral shape change little; the hadronic part remains prevalent in an 
extended region around the $\rho$ mass at all collision energies. The very 
fact that an appreciable reshuffling of hadronic and QGP contributions 
from the transition region occurs, indicates that the latter is a dominant 
source of low-mass dileptons at both RHIC and LHC. This creates a favorable 
set-up for in-depth studies of the chiral restoration transition in a 
regime of the phase diagram where quantitative support from lQCD computations 
for order parameters and the EM correlator is available.
%The extension of existing calculations to finite 3-momentum, and 
%to below $T_{\rm pc}$ can be expected for the midterm future. 
Current ideas of how to render these connections rigorous have been 
reported elsewhere. 
Phenomenologically, it turns out that current RHIC data for LMR dilepton 
spectra are consistent with the melting $\rho$ scenario\footnote{With
the caveat of the central Au-Au PHENIX data.}, including a recent first 
measurement of an excitation function all the way down to SPS energies. 
However, the accuracy of the current data does not yet suffice to 
discriminate in-medium spectral functions which differ considerably in
detail. These ``details" will have to be ironed out to enable definite 
tests of chiral restoration through the EM spectral function. 

While the low-mass shape of the spectra is expected to be remarkably 
stable with collision energy, large variations are predicted in the 
excitation function of other dilepton observables. First, the total 
yields increase substantially with collision energy. In the LMR, the 
dependence on charged-particle rapidity density, $N_{\rm ch}^\alpha$, 
is estimated to scale as $\alpha\simeq1.8$ from RHIC to LHC, significantly
stronger than as function of centrality at fixed $\sqrt{s}$. This is,
of course, a direct consequence of the longer time  it takes for the 
fireball to cool down to thermal freezeout. For the RHIC beam-energy scan
program, this opens an exciting possibility to search for a non-monotonous
behavior in the fireball's lifetime due to a critical slowing down of
the system's expansion. If the LMR radiation indeed emanates largely   
from the transition region, a slowed expansion around $T_c$ would take
maximal advantage of this, thus rendering an ``anomalous dilepton 
enhancement" a promising signature of a critical point.      
 
Second, the transverse-momentum spectra of thermal dileptons are expected 
to become much harder with collision energy, reflecting the increase in 
the collective expansion generated by the QGP prior to the transition 
region. This ``QGP barometer" provides a higher sensitivity than final-state 
hadron spectra which include the full collectivity of the hadronic evolution. 
The inverse-slope parameters for $q_t$ spectra in the LMR are expected to 
increase from $\sim$220\,MeV at SPS, to $\sim$280\,MeV at RHIC-200, and up 
to $\sim$450\,MeV at LHC-5500. Even larger values are reached in the IMR, 
although the situation is a bit more involved here, since 
(a) the QGP emission is increasingly emitted from earlier phases, and
(b) the hadronic emission, while picking up the full effect of additional 
collectivity at $T_{\rm pc}$, becomes subleading relative to the QGP. The 
trend in the LMR seems to line up with the recent slope measurements in 
photon excess spectra at RHIC and LHC. A similar connection exists for
the elliptic flow; pertinent data will be of great interest.   
Invariant-mass spectra in the IMR remain the most promising observable to 
measure early QGP temperatures, once the the correlated heavy-flavor decays 
can be either subtracted or reliably evaluated theoretically.

The versatility of dileptons at collider energies comprises a broad
range of topics, ranging from chiral restoration to direct-temperature
measurements, QGP collectivity, and fireball lifetime. Experimental
efforts are well underway to exploit these, while sustained theoretical 
efforts will be required to provide thorough interpretations.

%%%%%%%%%%%%%%%%%%%%%%%%%%%%%%%%%%%%%%%%%%%%%%%%
%% BACKMATTER
%%%%%%%%%%%%%%%%%%%%%%%%%%%%%%%%%%%%%%%%%%%%%%%%

\begin{theacknowledgments}
I gratefully acknowledge the contributions of my collaborators, 
in particular Charles Gale, Min He and Hendrik van Hees. 
I also thank the STAR and ALICE collaboration for making available 
their plots shown in this article. 
This work has been supported by the U.S. National Science Foundation
under grant nos.~PHY-0969394 and PHY-1306359, and by the A.-v.-Humboldt 
Foundation (Germany). 
\end{theacknowledgments}

%%%%%%%%%%%%%%%%%%%%%%%%%%%%%%%%%%%%%%%%%%%%%%%%
%% For The AIP proceedings layouts use either
%%%%%%%%%%%%%%%%%%%%%%%%%%%%%%%%%%%%%%%%%%%%

\bibliographystyle{aipproc}   % if natbib is available
%\bibliographystyle{aipprocl} % if natbib is missing

%%%%%%%%%%%%%%%%%%%%%%%%%%%%%%%%%%%%%%%%%%%%%%%%%%%%%

\end{document}